# Unlocking Ultrastrong High-Temperature Ceramics: Beyond Equimolar Compositions in High-Entropy Nitrides

O.V. Pshyk[a,b,c,d*], A. Vasylenko[e], P. Küttel[f], B. Wicher[b,g], P. Schweizer[a], J. Michler[a], T.E.J. Edwards[a*]

*a* Empa — Swiss Federal Laboratories for Materials Science and Technology, Laboratory for Mechanics of Materials and Nanostructures, Thun 3602, Switzerland

*b* Thin Film Physics Division, Department of Physics (IFM), Linkoping University, 581 83 Linkoping, Sweden

*c* NanoBioMedical Centre, Adam Mickiewicz University, Poznań, Poland

*d* Empa — Swiss Federal Laboratories for Materials Science and Technology, Laboratory for Surface Science and Coating Technology, Dübendorf 8600, Switzerland

*e* Department of Chemistry, University of Liverpool, Liverpool, United Kingdom

*f* Alemnis AG, Schorenstrasse 39, 3645 Gwatt, Switzerland

*g* Faculty of Materials Science and Engineering, Warsaw University of Technology, 02-507 Warsaw, Poland

*Corresponding authors: oleksandr.pshyk@empa.ch, thomas.edwards@empa.ch

Traditionally, increasing compositional complexity and chemical diversity of high-entropy ceramics whilst maintaining a stable single-phase solid solution has been a primary design strategy for developing new ceramics. Here, we unveil a groundbreaking alternative strategy based on deviation from conventional equimolar composition towards non-equimolar composition space, enabling tuning the metastability level of the supersaturated single-phase solid solution. By employing high-temperature micromechanical testing of refractory metal-based high-entropy nitrides, we found that the activation of an additional strengthening mechanism upon metastable phase decomposition propels the yield strength of a non-equimolar nitride at 1000 °C to a staggering 6.9 GPa, that is 40% higher than the most robust equimolar nitride. We show that the inherent instability triggers the decomposition of the solid solution with non-equimolar composition at high temperatures, inducing strengthening due to the coherency stress of a spinodally modulated structure, combined with the lattice resistance of the product solid solution phase.

# Introduction

Refractory metal-based nitrides belong to the broad family of refractory ceramic materials (a set of carbides, nitrides, and borides of the group IV and V transition metals) because of their ultra-high melting temperatures ($T_m$ above ~3000 °C) and phase stability they represent the only suitable class of materials available for a diverse range of mechanically loaded applications where components or tools are subjected to the most extreme of operating environments[1]. Delving into high-entropy ceramics[2], the number of possibilities for exploring refractory nitrides is much greater. In such multi-component systems composed of at least five principal unary ceramic compounds, a high configurational entropy (set here by a disordered multi-cation sublattice – assuming that the anion sublattice remains intact) is expected to produce a preference for single-phase solid solutions with simple crystal structures. The presence of many chemically distinct atoms on the metal sublattice produces attractive effects, such as sluggish diffusion and considerable lattice distortion, amongst others[3], leading to properties often highly surpassing those of the constituent conventional ceramic components[4]. However, although many studies focus on maximizing configurational entropy (using equimolar ratios of metal elements), indeed successfully achieving stabilization of single-phase solid solution in high-entropy oxides[5], the impact of configurational entropy on the solid solution phase stability may have been overestimated in metallic alloys[6,7] and nitrides[8,9] – configurational entropy may not even be a vital parameter for the design of multi-component alloys and ceramics with superior properties[10–12] (Supplementary Note 1). The vast composition space of metastable non-equiatomic high-entropy ceramics instead provides even more ways for exploration of new ceramics featuring complex phase transformation sequences upon metastable phase decomposition[13].

High-temperature mechanical properties are crucial for the operation of ceramics in extreme environments[14], yet have been little studied thus far in high-entropy ceramics[15]. Although the unique mechanical properties of high-entropy ceramics are foremost determined by solid solution strengthening, among other beneficial mechanisms (see Supplementary Note 2), less attention has been placed on the stability of this mechanism at high temperatures, limiting the potential impact of these materials in applications. Inherent metastability of multi-component nitrides[8], in turn, can trigger different strengthening mechanisms upon single-phase solid solution decomposition, particularly when exposed to mechanical and thermal loads, leading to their higher strength and damage tolerance. The remarkable improvement of mechanical properties of conventional nitrides due to the formation of three-dimensional nanostructure[16] or precipitation of two-dimensional atomic-plane-thick Guinier-Preston (GP) zones[17] upon decomposition of single-phase solid solutions are proofs of the concept of metastability tuning in high-entropy nitrides.

Here, we demonstrate that tuning the metastability of high-entropy nitride ceramic thin films by deviating from equimolar metal ratios toward non-equimolar composition space allows for decreasing the stability of the solid solution, thus enabling its decomposition at elevated temperature. This entails two key benefits: coherency strain strengthening due to dual-phase isostructural nanodomains formed upon spinodal decomposition of the solid solution and retained solid solution hardening of the high-entropy nitride phase. This combination leads to good retention of high strength at elevated temperatures. At the same time, we show that maximization of configurational entropy and chemical diversity in equimolar high-entropy nitrides does not maintain high strength at elevated temperatures. Due to the relatively greater stability of all equimolar high-entropy nitrides at high temperatures, strength-inducing phase transformations are absent or insufficient to retain high yield strength in the studied temperature range.

# Results

Refractory-metal-based pentanary (TiHfNbVZr)N, hexanary (TiHfNbVZrTa)N, and heptanary (TiHfNbVZrTaW)N thin films, 6 μm thick, are grown by direct current magnetron sputtering (DCMS) in an industrial deposition system on (0001) single crystalline sapphire substrates; low and high additions of Al to the pentanary system are enabled by a hybrid co-sputtering process based on high power impulse magnetron sputtering (HiPIMS) and DCMS. The fraction of metal elements on the cation sublattice was determined by X-ray photoelectron spectroscopy (XPS) and time-of-flight elastic recoil detection analysis (ToF-ERDA): pentanary $(Ti_{0.23}Hf_{0.15}Nb_{0.17}V_{0.21}Zr_{0.24})N$, hexanary $(Ti_{0.23}Hf_{0.15}Nb_{0.13}V_{0.14}Zr_{0.21}Ta_{0.14})N$, heptanary $(Ti_{0.17}Hf_{0.11}Nb_{0.14}V_{0.14}Zr_{0.16}Ta_{0.11}W_{0.17})N$, hexanary 'Al-low' $(Ti_{0.21}Hf_{0.13}Nb_{0.14}V_{0.18}Zr_{0.20}Al_{0.14})N$ and hexanary 'Al-high' $(Ti_{0.13}Hf_{0.07}Nb_{0.07}V_{0.10}Zr_{0.12}Al_{0.51})N$, see Supplementary Table S1. Similarly, the anion sublattice is almost fully occupied with N, with a low amount of vacancies; more details about the microstructure and elemental composition of the as-deposited films are given in our previous report[11].

XRD patterns (Fig. 1a, b, Fig. S1) indicate all as-deposited films are single-phase B1-NaCl-structured solid solutions, exhibiting dense fine-fibrous microstructures with column diameters of 80-100 nm and mixed polycrystalline orientations (demonstrated in our previous

report[11]), typical for refractory-metal-based nitrides. In short, we have synthesized single-phase high-entropy nitrides with near-equimolar, and non-equimolar, metal atom ratios. All nitrides with equimolar metal ratios retain the single phase structure upon annealing up to 1200 °C, demonstrating XRD patterns with peaks corresponding to the B1-NaCl-type phase structure, with low-intensity broad shoulders to (111) and (200) peaks (Fig. S1). Annealing at any higher temperature results in the oxidation of the thin film surface (as evident for heptanary thin films after annealing at 1200 °C), making assessing high-entropy nitride hardness by nanoindentation impossible. However, the hexanary Al-high single-phase solid solution undergoes decomposition at 900 °C already, evident as strong left- and right-hand shoulders to the (111) and (200) peaks, see Fig. 1a; this indicates the formation of additional B1-NaCl phases with a slight difference in lattice parameters compared to the original (matrix) phase. XRD patterns showing this for two critical representative compositions are shown in Fig. 1a, b, while the remainder of the cases are given in Fig. S1. Density-functional theory (DFT) calculations of formation energy for the pentanary system and hexanary system with different fractions of Al on the metal sublattice predicts all compositions to be metastable against decomposition to binary or ternary compounds (Fig.1c). Importantly, the driving force to decomposition increases as a function of Al fraction. All above confirms the targeted stability trend can be realized by designing a specific composition of the nitrides, which is in agreement with DFT calculations – without yet considering the kinetics of the decomposition process that can be significantly altered by Al content or upon the onset of decomposition. An evaluation of mechanical properties of the films after annealing by nanoindentation, Fig. 1d, reveals good retention of as-deposited hardness up to at least 1100 °C in all systems, with a minor (<10%) improvement in hardness seen in the refractory pentanary and hexanary nitrides. However, the Al-high and heptanary systems showed even greater improvement with highest hardness after annealing at 1000 °C (42.6±2.3 GPa) and 1100°C (44.6±1.9 GPa), respectively. Scanning transmission electron microscopy (STEM) of the highest hardness conditions is carried out to understand the microstructural evolution generating such performance improvements – the as-deposited conditions have been previously characterized and reported in[11]. Whilst the pentanary nitride annealed at 1000 °C displays an unchanged columnar microstructure, Fig. S2a, local chemical segregation across the metal sublattice is evidenced by atomic resolution high-angle annular dark field (HAADF) imaging, Fig. 2a. The heptanary nitride annealed at 1100 °C exhibits grain-scale segregation, in particular of W (Fig. S2b, Fig. S4), a considerable density of twin boundaries, Fig. 2b, as well as atomic-scale segregation inside the grain, Fig. 2b insert, but no evidence of W-rich Guinier Preston zones previously reported for W-doped ternary nitride (Ti,Al)N[17]. Finally, both Al-containing nitrides displayed significant nano-scale chemical segregation after annealing at 1200 °C, Fig. 2c-f and Fig. S5, despite the selected area electron diffraction (SAED) patterns showing only reflections matching the B1-NaCl structure phase (Fig. S6). EDS mapping (Fig. S5) and EDS line scans (Fig. S7, Fig. 2d, f inserts) reveal the separation of all refractory metals from Al between the two contrasting isostructural domains with a modulation wavelength of the order 3.5 ± 0.4 nm for hexanary 'Al-low' and 12.2 ± 2.2 nm hexanary 'Al-high' films. Such decomposition of supersaturated solid solution into coherent cubic nanodomains via spinodal mechanism is typical for Al-containing refractory metal nitrides[16]. The non-equimolar solid solution decomposes at elevated temperature due to the high thermodynamic driving force to decomposition (Fig. 1c) while all equimolar solid solutions are relatively more stable (see XRD above, Fig. 1a, b, Fig. S1). To the best of our knowledge, this is the first evidence of such a decomposition in high-entropy nitrides, including in the Al-low system where the thermodynamic driving force for decomposition is the lowest.

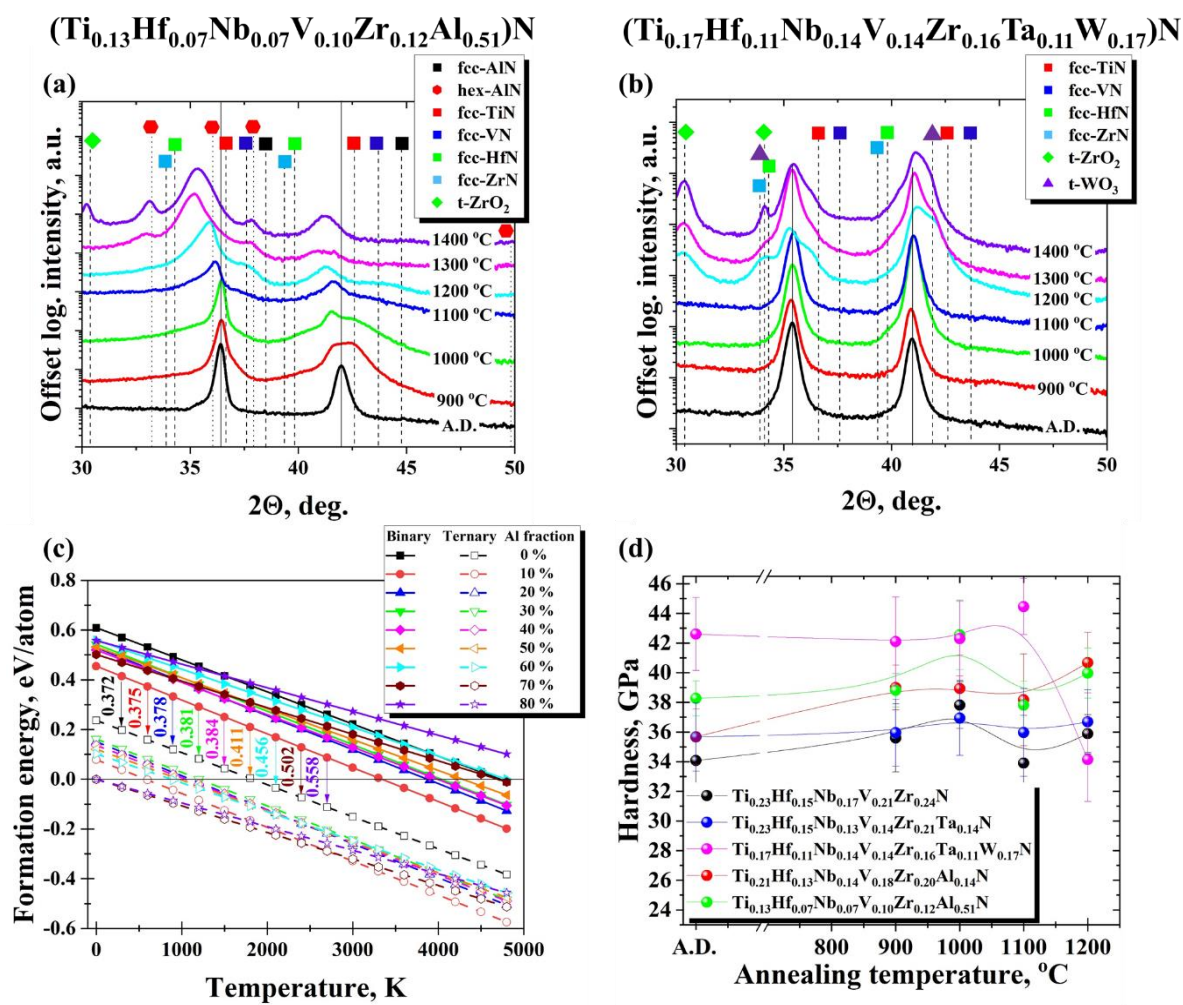

Fig. 1. XRD patterns for (a) hexanary 'Al-high' $(Ti_{0.13}Hf_{0.07}Nb_{0.07}V_{0.10}Zr_{0.12}Al_{0.51})N$ and (b) heptanary $(Ti_{0.17}Hf_{0.11}Nb_{0.14}V_{0.14}Zr_{0.16}Ta_{0.11}W_{0.17})N$ thin films before and after annealing in vacuum at 900, 1000, 1100, 1200, 1300, 1400 °C. (c) Calculated formation energies, $E_f$, of B1-NaCl structured pentanary system and hexanary system with different Al fractions as a function of temperature. $E_f$ for systems with metal ratios on the cation sublattice similar to experimental samples is calculated considering binary (solid lines, full symbols) or ternary (dashed line, empty symbols) decomposition products. The decomposition potential, i.e. driving force for decomposition, is determined as the energy difference between these formation energies (arrows with rotated text give this decomposition potential for each composition in eV/atom, at 0 K). The thermodynamic driving force for decomposition increases with the increase of Al content in the solid solution. This trend suggests that the solid solutions with higher Al content possess the highest driving force for the decomposition into less complex phases or solid solutions, even at high temperatures. (d) Nanoindentation hardness of all films performed before and after annealing in a vacuum at different temperatures of 900-1200 °C.

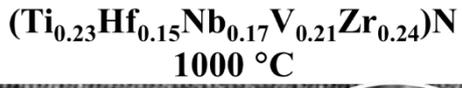
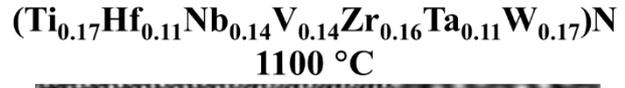
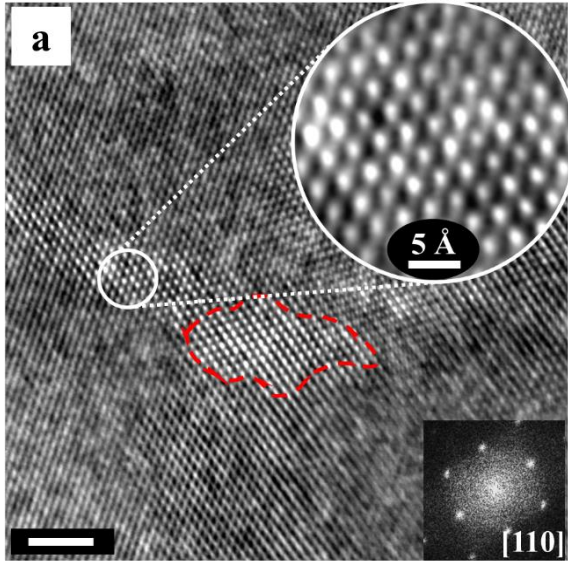
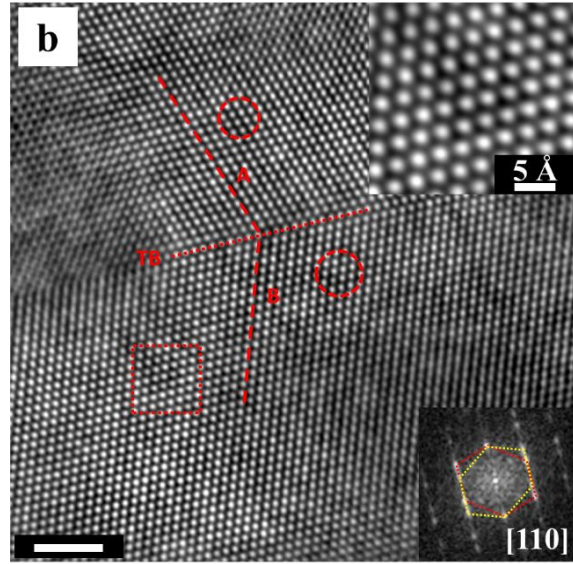
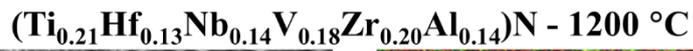
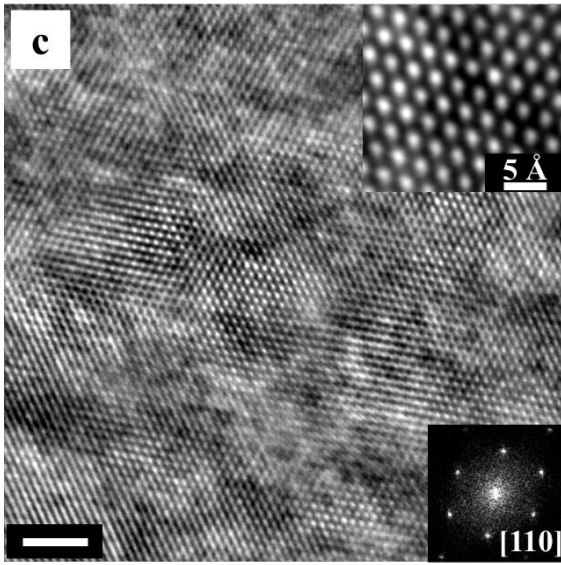
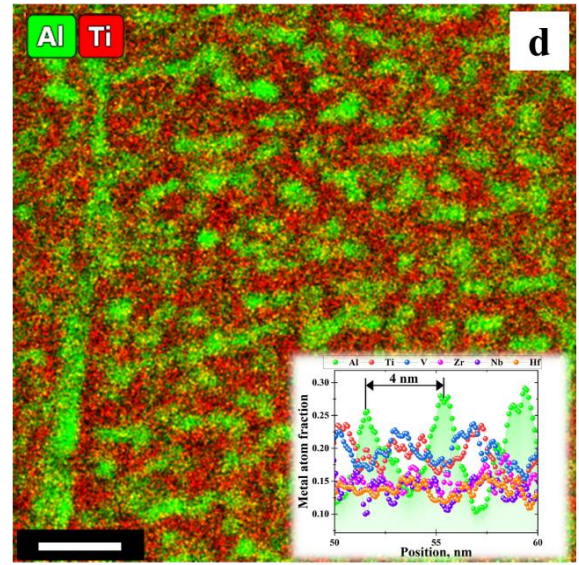
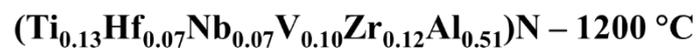
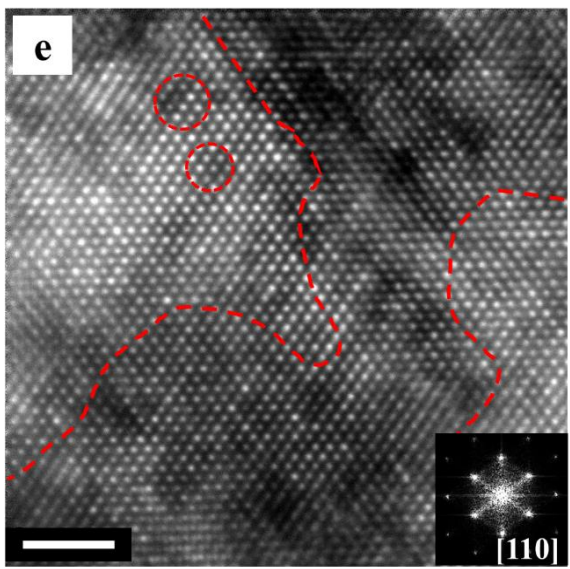
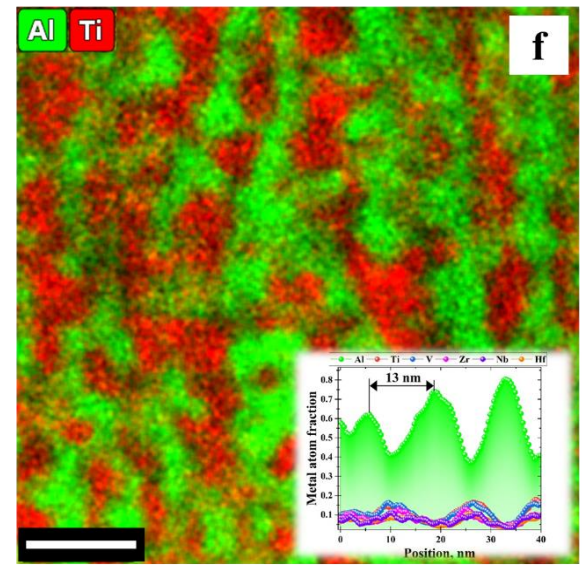

Fig. 2. (a) Cross-sectional STEM-HAADF acquired along [110] zone axis from pentanary nitride annealed at 1000 °C and (b) heptanary nitride annealed at 1100 °C. Cross-sectional STEM-HAADF images taken along [110] zone axis and STEM-EDX color-coded combined maps of Ti-Al acquired from (c, d) the hexanary 'Al-low' and (e, f) hexanary 'Al-high' films after annealing at 1200 °C. Red dashed lines show the borders of the high contrast (high atomic number, Z) areas, and red dashed circles mark the local chemical segregation across the metal sublattice enlarged in the upper right insets in (a), (b), and (c). Fast Fourier transform (FFT) patterns of (a), (b), (c), and (e) are presented in bottom right inserts. Scale bars in (a) – (c) and (e) are 2 nm; in (d) it is 10 nm and in (f) – 20 nm.

High-temperature microcompression is carried out to capture mechanical performance and assess the effectiveness of nitride microstructure strategies in conditions approximating extreme applications. Micro-pillars ~1.8 μm in diameter, Fig. S8, produced by focused ion beam (FIB) milling, where the height is less than the coating thickness, are compressed *in vacuo* at six temperatures, changing from the room temperature (~ 25 °C) to 1000 °C, using a flat conductive diamond punch. Compressive yield strengths using a 1% strain criterion, $\sigma_y$, Fig. 3a, indicate initially greater room temperature strength of the pentanary and hexanary Al-high systems ($\sigma_y$ ~ 11–12 GPa), while all other equimolar systems have $\sigma_y$ ~ 8–9 GPa. Nevertheless, all such pillars fail by catastrophic brittle fracture; this is seen in stress-strain curves, Fig. 3b, and *post-mortem* SEM images, see Fig. S9, which also include all other test temperatures. In the 500 to 800 °C range, all systems are softer than at 25 °C, yet broadly show strength stability across this range, with the pentanary nitride showing the most significant strength loss from 25 °C, to match the other equimolar systems in this temperature range closely. Ductility, through cohesive plasticity, over several percent strain, is now observed before eventual cracking-induced rapid strength loss. However, the hexanary Al-high system has at least 32% higher $\sigma_y$ in this range. From 800 to 1000 °C, the best performing system among equimolar compositions is the pentanary high-entropy nitride, whilst the heptanary and hexanary Al-low systems show the most rapid decrease of $\sigma_y$ with temperature. Again, the non-equimolar Al-high system demonstrates 42-43% higher $\sigma_y$ in this range than the best-performing equimolar system, with $\sigma_y$ = 6.9 ± 0.5 GPa at 1000 °C. Deformation above 900 °C is predominantly cohesive, with local buckling of the upper sections of the pillars or shear plane formation (see TEM below).

It can be seen from the exemplary microcompression loading curves for hexanary 'Al-high' pillars in Fig. 3b and the complete set of material systems and test temperatures in Fig. S10, that there is a noticeable reduction in measured elastic loading modulus, from either 800 or 900 °C depending on composition, with 1000 °C showing the lowest modulus. This could be associated with chemical reactivity between pillar and punch producing additional interface reaction phases, as well as plastic deformation of the diamond itself, as detailed in Supplementary Note 3. Despite this, we consider the yield points of the loading curves to remain sufficiently distinctive for useful interpretation.

TEM imaging and analysis following microcompression at 1000 °C revealed the spinodally decomposed microstructure in the Al-rich hexanary nitride is retained as previously in the annealed film, Fig. 4, with a modulation wavelength ~10 nm, and preferential alignment of the Al-rich nanodomains along the film growth direction – i.e., the [001] axis, considering the predominant (002) film texture here. The bending-over of the Al-rich and refractory-metal-rich nanodomains is evident from localized plastic buckling upon compression in the upper half of the pillar (Fig. 4a, b, c). Yet, the pillar remains dense: no evidence of cracking or decohesion of the columnar grains at boundaries is seen. Despite the high vacuum environment, multiple oxidation layers surround the Al-high hexanary nitride pillar compressed at 1000 °C: an inner $(Ti, V)O_x$, an intermediate $ZrO_x$, and finally, an outer $AlO_x$ (Fig. S18, Fig. S19). The $AlO_x$ outer oxide layer acts as an oxygen diffusion barrier, preventing oxygen inward diffusion[18], hence avoiding grain boundary oxide decoration in the pillar bulk. In stark contrast, extensive oxygen inward diffusion is observed for the W-containing heptanary nitride, see Supplementary Note 4.

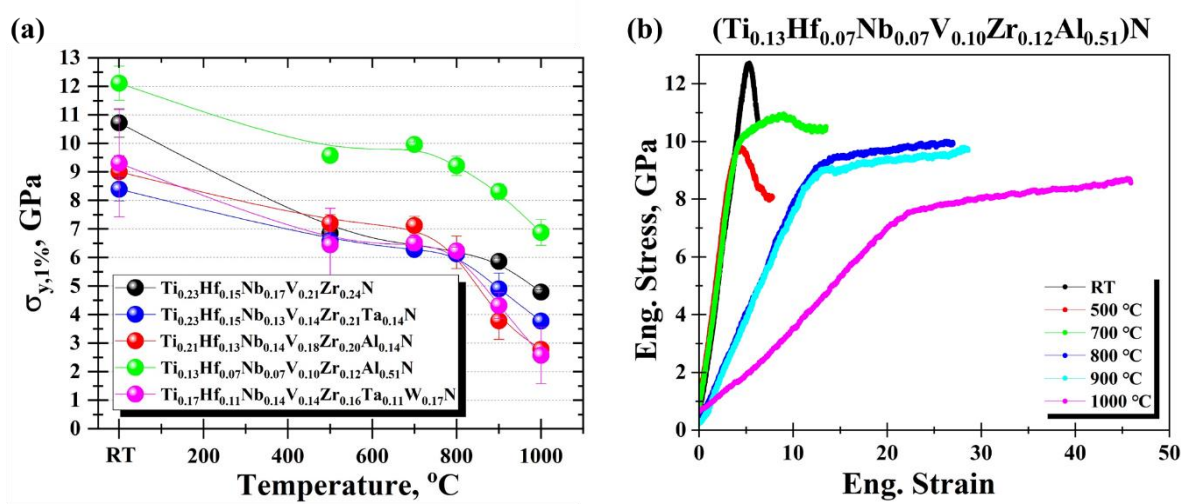

Fig. 3. (a) Yield strength at 1% strain measured at different temperatures and (b) corresponding representative stress-strain curves for Al-high pillars compressed at RT, 500 °C, 700 °C, 800 °C, 900 °C, and 1000 °C. For statistical accuracy, at least 4-5 pillars are compressed for each film composition at each test temperature, but only representative stress-strain curves are shown here.

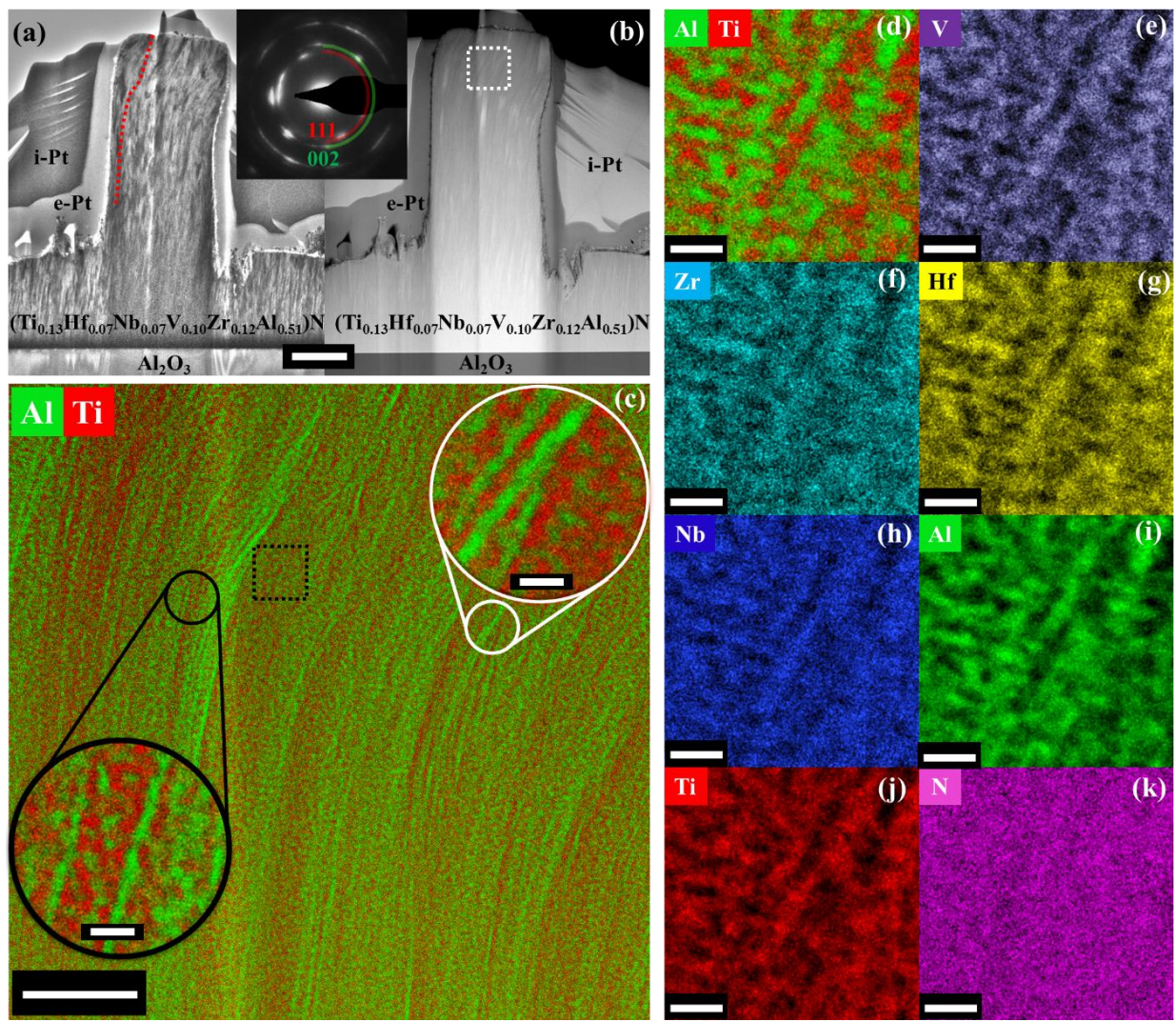

Fig. 4. Overview cross-sectional (a) BF-TEM (red dotted line serves as a guide to buckling of the columnar grains) and (b) Z-contrast STEM images from the hexanary 'Al-high' pillar after compression at 1000 °C with corresponding SAED pattern as an inset, and electron and ion-deposited Pt protective straps (e-Pt & i-Pt, respect.) indicated. (c) EDX color-coded maps of Al and Ti acquired from the white square in (b) with insets demonstrating EDS maps with a higher magnification. EDS color-coded maps of (d) Ti-Al, (e) V, (f) Zr, (g) Hf, (h) Nb, (i) Al, (j) Ti, and (k) N. Scale bar in (a) and (b) is 1µm, in (c) it is 200 nm (20 nm in the insets), and in (d) – (k) they are 20 nm.

## Discussion

Deformation at room temperature remains brittle for all high-entropy nitride systems, but with a higher $\sigma_y$: lacking in effective dislocation plasticity because highly directional covalent bonding, the majority constituent of such nitrides[19], leads to a high lattice resistance[20]. The high-entropy nitrides display initially ductile plasticity from 500 °C, which is consistent with expectation that thermal activation of bending and turning of interatomic bonds and annihilation of ion-induced point defects decrease the lattice resistance (Peierls stress) and activate the motion of dislocations in the 0.1-0.3 $T_m$ range[14]. All test temperatures are within the latter range, considering the melting point of the studied high-entropy nitrides can be in the 3200-3400 K[21] range if predicted by the rule-of-mixture. Commonly in B1-NaCl structured refractory nitrides, thermalisation of electrons into less localised and less directional metallic states leads to the activation of additional slip systems, now occurring also on {111}<110>, in addition to {110}<110>, thereby inducing additional ductility[22,23]. The microcompression trends in ductility and $\sigma_y$ observed here are consistent with previous high-temperature mechanics work[24] on simpler B1-NaCl structured multi-component nitrides, where characterisation was limited to 500 °C – the exact compositions, equimolar or otherwise, not being reported therein hindering fruitful comparison.

Amongst the equimolar compositions, the pentanary system $(Ti_{0.23}Hf_{0.15}Nb_{0.17}V_{0.21}Zr_{0.24})N$ is the strongest in microcompression over almost the entire temperature range from 25 to 1000 °C, particularly at the extremes. Strengthening in all equimolar cases can be attributed to the lattice resistance of the solid solution – consistently improved neither by greater chemical complexity nor by the addition of higher melting point nitride formers. The sluggish diffusion characteristic of multi-principle component alloys or high-entropy metal sub-lattice nitrides[25], combined with a relatively low driving force for decomposition, means equimolar systems retain their single-phase structure. Hence, $\sigma_y$ decreases at high temperatures, which is primarily associated with the decline of the lattice resistance and absence of other obstacles for dislocation motion[14]. An essential feature of the annealing study is that hardness increases in all the analyzed systems upon annealing above the deposition temperature (450 °C); this is surprising, as commonly hard coatings without an additional microstructural strengthening mechanism would hence lose the strength benefit of a high density of ion-bombardment-induced crystal defects formed during synthesis[26]. As no modification to the 100 nm-scale columnar grain structure upon annealing is evident, a potential strengthening effect needs to be attributed to chemical segregations at the atomic-scale and/or short-range ordering observed in the metal sub-lattice, Fig. 2a, which may alter local lattice resistance.

The case of the W-containing heptanary nitride is particularly thought-provoking, as the high performance of the annealing study is not borne out in microcompression. The Tabor factor, that is, hardness to yield strength ratio, of 4.6 for the heptanary nitride is indicative of brittleness-limited performance[24] – compared to the pentanary and hexanary Al-high nitrides, where cohesive plasticity carries a theoretical Tabor factor of ~3 [27]. It is also important to note the considerably longer thermal holds at each test temperature experienced by the thin films during microcompression testing, lasting 4 h at least, Fig. S13, as compared with 10 min for the annealing study, which justifies oxidation-induced strength loss. Moreover, the prolonged time for diffusion in microcompression tests may favor annihilation of the ion-bombardment-induced point defects, which results in weakening of this strengthening effect, starting from 500 °C. On the other hand, the cause of hardness improvement between annealing at 1000 and 1100 °C remains unclear – only grain-scale W segregations (Fig. S3b, Fig. S4) and short-range ordering, Fig. 2b, are identified here. Surprisingly, the presence of twin boundaries distinguishes the W-containing heptanary system from the other equimolar compositions; these are uncommon for refractory nitrides due to high stacking fault energies[23] and may contribute to high hardness. Since the system contains refractory metal nitrides with low (NbN, VN, and TaN) and high (HfN, TiN, ZrN) stacking fault energies[23], the formation of twins in high-entropy nitrides is energetically favoured[28], playing therefore a hardening role especially when twin thickness becomes larger at the elevated temperatures[29]. Combined with W segregations and short-range ordering, such twin boundaries may be further exploitable if better understood. Moreover, any possible departures from ideal randomness, i.e. short-range ordering or clustering, at high annealing temperature can significantly alter the Peierls stress and related strengthening[30].

The greatest relative hardness increase is, however, achieved by the hexanary Al-high system $(Ti_{0.13}Hf_{0.07}Nb_{0.07}V_{0.10}Zr_{0.12}Al_{0.51})N$, which undergoes microstructural decomposition most rapidly – consistent with the higher thermodynamic driving force (Fig. 1a, c) compared with equimolar compositions. The observed XRD peak shoulders, compositional modulation, and retention of B1-NaCl structure are consistent with spinodal-type decomposition typical for Al-rich refractory metal nitrides[31]. Decomposition of the Al-high solid solution sub-lattice into isostructural (TiHfVNbZr)N- and AlN-rich coherent nanodomains, Fig. 5a, results in spatially fluctuating strain fields between the domains due to the negligible difference in their lattice parameters[32]. These coherency strains can also be accommodated by partial dislocations, stacking faults, and misfit dislocation depending on the stage of decomposition[33,34]. Hence, its here-unmatched high temperature strength is provided by coherency strain between decomposed nano-domains[35] as well as solid solution strengthening of the spinodal phase with a composition part-way between the pentanary and hexanary Al-low nitrides. Stacking faults and partial dislocations should enable good ductility, and activate different slip systems, although the results presented here indicate that the pentanary nitride itself nevertheless presents equally cohesive room temperature plasticity. Preferential alignment of the spinodal dual-skeleton of percolating phases likely leads to additional strength retention along the growth direction at high temperature[36]. The hexanary 'Al-low' system also decomposes through this mechanism after annealing at 1200 °C, but the high-temperature strength up to 1000 °C of this system is in the range of all other equimolar systems – likely due to delayed onset of decomposition of the hexanary 'Al-low' compared to the 'Al-high' system, given the lower thermodynamic driving force (Fig. 1c).

Moreover, the high-temperature strength of the Al-high system can be explained by a larger amplitude of the composition modulation[35] compared to the Al-low system (Fig. 2d, f, Fig. S7). The hexanary 'Al-low' system, annealed 1200 °C, shows the highest hardness (Fig. 1d), implying that the transformation of c-AlN nanodomains into thermodynamically preferred yet softer hexagonal AlN is delayed in the Al-low system in comparison to the Al-high system[8,37] as evident from XRD results (hexagonal AlN phase precipitation starts at 1300°C in 'Al-high' films, Fig. 1a, and at 1400 °C in 'Al-low' films, Fig. S1c). Such a trend in the evolution of microstructure and phase composition of the 'Al-low' system may provide a more beneficial strengthening effect as compared to the 'Al-high' condition, at temperatures above 1100 °C, Fig. 5a. A classical summative strengthening model of the shear modulus-normalized shear strength, $\tau/G$, plotted component-wise in Fig. 5b for 'Al-high' system, features theoretical contributions from the solution-strengthened lattice, dislocation arrays, grain and test-piece sizes, coherency strains due to decomposition, as well as the less-understood short-range ordering[38,39], all detailed further in Supplementary Note 5. The opportunity to offset thermal softening by forming decomposed isostructural phases with higher compositional segregation at a greater spatial length scale is demonstrated in Fig. 5b, which may be achievable through a tailored composition design strategy that aims for the optimal synthesis of metastable phases highly prone to decomposition at specific application temperatures.

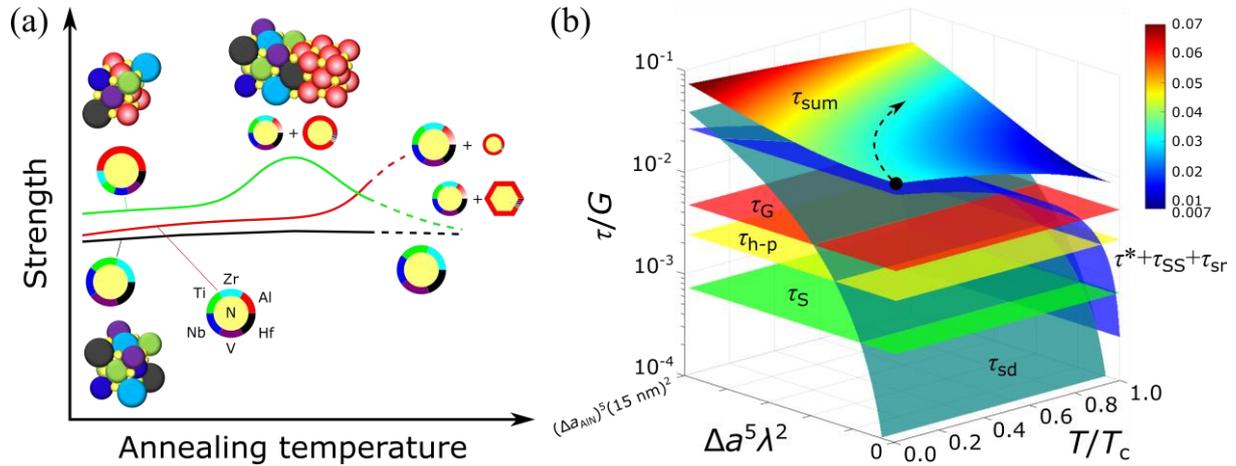

Fig. 5. Metastability tuning to optimize microstructural strengthening mechanisms of high-entropy nitrides for application at specific temperatures. (a) The schematics of strength evolution for all systems studied as a function of annealing temperature and related relative changes in constituent elemental composition and phase fractions. The black solid line represents the behavior of all equimolar refractory metal-based high-entropy nitrides, the red solid line represents the hexanary 'Al-low' system, and the green line stands for the hexanary 'Al-high' system. In (a), areas of circles represent the elemental compositions and phase proportions, where the fcc-NaCl structured cubic phase is denoted by circles and the hexagonal AlN phase is denoted by hexagons. The dashed lines show further hypothetical evolution of the strength and phase composition after annealing at temperatures beyond 1200 °C. Contributions to total shear strength $\tau_{sum}$ depicted in (b) for the hexanary 'Al-high' system are: lattice resistance $\tau^*$, solid solution strengthening of the metallic sub-lattice, $\tau_{SS}$, Taylor (dislocation) hardening, $\tau_G$, pillar-size strengthening, $\tau_S$, grain-size strengthening, $\tau_{h-p}$, short-range ordering, $\tau_{sr}$, and spinodal decomposition, $\tau_{sd}$. $\Delta a^5 \lambda^2$ controls spinodal strengthening based on the lattice parameter change $\Delta a$ across the decomposed phases with spatial wavelength, $\lambda$, whereby the axis limit here is the extreme of pure fcc-NaCl structured AlN grew to $\lambda = 15$ nm. The arrow in (b) highlights a route for effective use of microstructure control by spinodal decomposition to offset the thermal softening of the lattice resistance.

The excellent high-temperature strength-ductility combination of the 'Al-high' hexanary system, due to a synergy of several hardening mechanisms accessible upon decomposition of the solid solution via a spinodal mechanism and sufficient to deform the diamond punch itself, reflects its serious contendership as a next-generation ultrastrong high-temperature ceramics. A simple Tabor factor of 3 predicts a hot hardness of ~20 GPa at 1000 °C. It also highlights the experimental difficulty in investigating such high-entropy ceramics – only $B_4C$ has been suggested as mechanically superior at high temperatures so far[40], yet it is unlikely to resolve the chemical reactivity issues.

In summary, the current findings show that rather than focusing on equimolar compositions and single-phase formation and stabilization in high-entropy ceramics, the unexplored realm of non-equimolar composition space can offer a pathway to impressively tune phase metastability and related complex phase transformation sequences upon metastable phase decomposition to achieve different strengthening mechanisms at high temperatures. Metastable phase decomposition of the high-entropy nitride studied here via a spinodal mechanism produces a chemically modulated microstructure generating coherency strains leading to an ultrahigh yield strength at 1000°C. Future technological developments in such high-entropy nitrides may aim to optimize the refractory metal chemistry for greater strength and microstructural stability beyond 1000 °C while retaining the benefits of the early decomposing solid solution bringing spinodally-modulated microstructure and oxidation resistance here and even exploit additional mechanisms such as high twin densities and atomic-scale segregations and re-arrangements. The strategy of metastability tuning demonstrated here for high-entropy nitrides may pave a route towards tailoring high-temperature mechanical properties in other high-entropy ceramics. We anticipate that a mesmerizing world of ultrastrong high-temperature ceramic materials can be unraveled by delving into the metamorphosis of metastable high-entropy nitride phases, employing, amongst others, spinodal decomposition.

# Methods

Refractory-metal-based pentanary (TiHfNbVZr)N, hexanary (TiHfNbVZrTa)N, and heptanary (TiHfNbVZrTaW)N thin films are grown by direct current (DC) magnetron sputtering in an industrial CemeCon AG CC800/9 magnetron sputtering system. The hexanary Al-low and Al-rich (TiHfNbVZrAl)N are grown by a hybrid high-power impulse and DC magnetron co-sputtering HiPIMS/DCMS, see previous publication for details[11]. The films are grown with 6 μm thickness on (0001) single crystalline sapphire substrates.

The micro-pillars are produced with the equivalent diameter of ~1.8 μm and the final taper angle not exceeding 3° by focused ion beam (FIB) milling in a Tescan Lyra 3 instrument. The micro-pillars are milled with heights of ~4-4.5 μm) to avoid sample/pillar penetration into the substrate during compression and to avoid pillar-on-pedestal effects.

Micro-pillar compression is performed *in situ* in a Zeiss DSM 962 scanning electron microscope (SEM, base pressure $1.3 \times 10^{-3}$ Pa) using an Alemnis SEM indenter extensively modified for operation at temperatures up to 1000 °C, with pyrolytic graphite-on-pyrolytic boron nitride heaters to achieve independent heating of both sample and indenter[41]. The compression is performed using a 5 μm diameter diamond flat punch manufactured for high-temperature experiments by Synton-MDP. To minimize the effect of thermal drift during compression at elevated temperatures, matching the sample temperature and indenter temperature follows the procedure described elsewhere[41] to ensure thermal equilibrium at contact, considering the impact of substantial mutual radiative heating above ~600 °C[42]. Initial tip temperature calibration is achieved by indentation of the molybdenum holder adjacent to a spot welded reference thermocouple. The calibration of the compliance of the loading frame is performed at test temperatures by indentation of the (0001) single crystalline sapphire substrate material using a diamond Berkovich indenter, according to the method in [43]. To achieve stabilization of the system at high temperatures, the heating rate is set to 5 °C/min. Prior to the test, the holding time for complete equilibration of tip/sample temperature and the rest of the loading frame and sensors is set to ~4-5 hours at each test temperature – see Supplementary Fig. S13 for an exemplary heating schedule. For statistical accuracy, at least 4-5 pillars are compressed for each film composition at each test temperature. The compression is conducted at room temperature (RT), 500 °C, 700 °C, 800 °C, 900 °C, and 1000 °C, once the isothermal conditions for the sample and punch are achieved. The micro-pillars are compressed in displacement control mode with a strain rate of $1 \times 10^{-3}$ s$^{-1}$ at temperatures from RT to 900 °C and with a strain rate of $1 \times 10^{-2}$ s$^{-1}$ at 1000 °C for practical reasons. The temperature-dependent compliance of the sapphire substrate[44,45] and the pillar geometry is considered by applying Sneddon's and Zhang's corrections[46] during yield strength calculations using the associated MicroMechanics Data Analyser (MMDA) software.

Film and micro-pillar cross-sectional lift-out for scanning and transmission electron microscopy (STEM) and selected area electron diffraction (SAED) are performed by FIB in a FEI Helios NanoLab G3 UC Dual Beam SEM/Ga$^+$ FIB system. TEM investigations are performed using a probe-corrected Thermo Fisher Scientific Titan Themis 200 G3 operated at 200 kV. Energy dispersive X-ray spectroscopy (EDX) is acquired with the integrated SuperX detector. The annealing is performed in a vacuum with the heating rate of 10 °C/min to the annealing temperature $T_a$, which is in the range 900-1200 °C, and then kept at $T_a$ for 10 min. Following anneals, the furnace is allowed to naturally cool to room temperature. The phase composition of the films before and after annealing is determined using Bragg-Brentano X-ray diffraction (XRD) using a PANalytical Empyrean X-ray diffractometer. To avoid reflections from the sapphire substrate, the XRD scans are recorded using a 1° offset along ω. The composition and stoichiometry of the films are determined by X-ray photoelectron spectroscopy (XPS) and time-of-flight elastic recoil detection analysis (ToF-ERDA). More details about the measurements and analyses are given in the reference[11].

The NanoTester Vantage Alpha by Micro Materials, Ltd. (Wrexham, Wales) equipped with a diamond Berkovich probe is used to measure nanoindentation hardness H for refractory-metal-based thin film nitrides. Depth sensing indentations are acquired with a constant load of 30 mN during a dwell period of 5 s, yielding indentation depths <10% of the film thicknesses. H values are calculated following the Oliver and Pharr[4] rule, considering the elastic unloading part of the generated load-displacement curve. The standard deviation errors are extracted from the corresponding 20 indents (each separated by 25 µm within the boundaries of square grids).

Energy above convex hull (formation energy $E_f$) for compositions $(TiHfNbVZr)_{1-x}Al_xN$ was calculated as

$$E_f(x) = E_{total} - \sum_i x_i E_{total}^i, \quad (S.1)$$

where $E_{total}$ is a total energy of a composition at concentrations $x$ modeled as quasi-randomly arranged atoms[47] in B1-NaCl structure, as computed with density-functional theory (DFT), $E_{total}^i$ is the total energy of potential decomposition products, with corresponding weightings $x_i$ to form a considered composition of interest –compositions in Ti-Hf-Nb-V-Zr-Al-N system, including binary and ternary compositions, respectively (as illustrated in Figure 1a), reported previously[48] – computed at the same level of approximations with DFT as implemented elsewhere[49,50]: kinetic energy cutoff $E_{cut}$ is 650 eV, k-spacing in Brillouin zone is 0.5, energy convergence threshold in self-consistent electronic calculations is $10^{-4}$ eV, ionic geometry relaxation is performed until the $10^{-3}$ eV threshold is reached.

To account for temperature effects a configurational entropy term is added to the expression $E_f$ in Eq. S.1:

$$E_f = E_f - TS \quad (S.2)$$

where

$$S = k_B \sum_i x_i \ln x_i \quad (S.3)$$

is configurational entropy, for calculation of which all atomic sites in B1-NaCl structure are considered equally accessible for all atomic species in compositions.

**Acknowledgements.** AVP acknowledges the financial support from the National Science Centre of Poland (SONATINA 2 project, UMO-2018/28/C/ ST5/00476), the Foundation for Polish Science (FNP) under START scholarship (START 72.2020), and Polish National

Agency for Academic Exchange under the Bekker Programme (PPN/BEK/2019/1/00146/U/00001). AV acknowledges the financial support from the National Science Centre of Poland (SONATINA 2 project, UMO-2018/28/C/ST5/00476). BW is grateful to the Polish National Agency for Academic Exchange for support upon the NAWA Bekker (BPN/BEK/2021/1/00366/U/00001) and the Warsaw University of Technology within the Excellence Initiative: Research University, IDUB Mobility program (1820/74/Z09/2023). TEJE acknowledges funding from the European Union's Horizon 2020 research and innovation programme under the Marie Skłodowska-Curie grant agreement No. 840222. The contributions of G. Greczynski and B. Bakhit are also gratefully acknowledged.

# Unlocking Ultrastrong High-Temperature Ceramics: Beyond Equimolar Compositions in High-Entropy Nitrides

## Supplementary Material

O.V. Pshyk[a,b,c,d], A. Vasylenko[e], P. Küttel[f], B. Wicher[b,g], P. Schweizer[a], J. Michler[a], T.E.J. Edwards[a]

*a Empa — Swiss Federal Laboratories for Materials Science and Technology, Laboratory for Mechanics of Materials and Nanostructures, Thun 3602, Switzerland*

*b Thin Film Physics Division, Department of Physics (IFM), Linkoping University, 581 83 Linkoping, Sweden*

*c NanoBioMedical Centre, Adam Mickiewicz University, Poznań, Poland*

*d Empa — Swiss Federal Laboratories for Materials Science and Technology, Laboratory for Surface Science and Coating Technology, Dübendorf 8600, Switzerland*

*e Department of Chemistry, University of Liverpool, Liverpool, United Kingdom*

*f Alemnis AG, Schorenstrasse 39, 3645 Gwatt, Switzerland*

*g Faculty of Materials Science and Engineering, Warsaw University of Technology, 02-507 Warsaw, Poland*

# Methods

Refractory-metal-based pentanary (TiHfNbVZr)N, hexanary (TiHfNbVZrTa)N, and heptanary (TiHfNbVZrTaW)N thin films are grown by direct current (DC) magnetron sputtering in an industrial CemeCon AG CC800/9 magnetron sputtering system. The hexanary Al-low and Al-rich (TiHfNbVZrAl)N are grown by a hybrid high-power impulse and DC magnetron co-sputtering HiPIMS/DCMS, see previous publication for details[1]. The films are grown with 6 μm thickness on (0001) single crystalline sapphire substrates.

The micro-pillars are produced with the equivalent diameter of ~1.8 μm and the final taper angle not exceeding 3° by focused ion beam (FIB) milling in a Tescan Lyra 3 instrument.

The micro-pillars are milled with heights of ~4-4.5 μm) to avoid sample/pillar penetration into the substrate during compression and to avoid pillar-on-pedestal effects.

Micro-pillar compression is performed *in situ* in a Zeiss DSM 962 scanning electron microscope (SEM, base pressure $1.3\times 10^{-3}$ Pa) using an Alemnis SEM indenter extensively modified for operation at temperatures up to 1000 °C, with pyrolytic graphite-on-pyrolytic boron nitride heaters to achieve independent heating of both sample and indenter[2]. The compression is performed using a 5 μm diameter diamond flat punch manufactured for high-temperature experiments by Synton-MDP. To minimize the effect of thermal drift during compression at elevated temperatures, matching the sample temperature and indenter temperature follows the procedure described elsewhere[2] to ensure thermal equilibrium at contact, considering the impact of substantial mutual radiative heating above ~600 °C[3]. Initial tip temperature calibration is achieved by indentation of the molybdenum holder adjacent to a spot welded reference thermocouple. The calibration of the compliance of the loading frame is performed at test temperatures by indentation of the (0001) single crystalline sapphire substrate material using a diamond Berkovich indenter, according to the method in [4]. To achieve stabilization of the system at high temperatures, the heating rate is set to 5 °C/min. Prior to the test, the holding time for complete equilibration of tip/sample temperature and the rest of the loading frame and sensors is set to ~4-5 hours at each test temperature – see Supplementary Fig. S13 for an exemplary heating schedule. For statistical accuracy, at least 4-5 pillars are compressed for each film composition at each test temperature. The compression is conducted at room temperature (RT), 500 °C, 700 °C, 800 °C, 900 °C, and 1000 °C, once the isothermal conditions for the sample and punch are achieved. The micro-pillars are compressed in displacement control mode with a strain rate of $1 \times 10^{-3}$ s$^{-1}$ at temperatures from RT to 900 °C and with a strain rate of $1 \times 10^{-2}$ s$^{-1}$ at 1000 °C for practical reasons. The temperature-dependent compliance of the sapphire substrate[5,6] and the pillar geometry is considered by applying Sneddon's and Zhang's corrections[7] during yield strength calculations using the associated MicroMechanics Data Analyser (MMDA) software.

Film and micro-pillar cross-sectional lift-out for scanning and transmission electron microscopy (STEM) and selected area electron diffraction (SAED) are performed by FIB in a FEI Helios NanoLab G3 UC Dual Beam SEM/Ga$^+$ FIB system. TEM investigations are performed using a probe-corrected Thermo Fisher Scientific Titan Themis 200 G3 operated at 200 kV. Energy dispersive X-ray spectroscopy (EDX) is acquired with the integrated SuperX detector. The annealing is performed in a vacuum with the heating rate of 10 °C/min to the annealing temperature $T_a$, which is in the range 900-1200 °C, and then kept at $T_a$ for 10 min. Following anneals, the furnace is allowed to naturally cool to room temperature. The phase composition of the films before and after annealing is determined using Bragg-Brentano X-ray diffraction (XRD) using a PANalytical Empyrean X-ray diffractometer. To avoid reflections from the sapphire substrate, the XRD scans are recorded using a 1° offset along ω. The composition and stoichiometry of the films are determined by X-ray photoelectron spectroscopy (XPS) and time-of-flight elastic recoil detection analysis (ToF-ERDA). More details about the measurements and analyses are given in the reference[1].

The NanoTester Vantage Alpha by Micro Materials, Ltd. (Wrexham, Wales) equipped with a diamond Berkovich probe is used to measure nanoindentation hardness H for refractory-metal-based thin film nitrides. Depth sensing indentations are acquired with a constant load of 30 mN during a dwell period of 5 s, yielding indentation depths <10 % of the film thicknesses. H values are calculated following the Oliver and Pharr[4] rule, considering the elastic unloading part of the generated load-displacement curve. The standard deviation errors are extracted from the corresponding 20 indents (each separated by 25 µm within the boundaries of square grids).

Energy above convex hull (formation energy $E_f$) for compositions (TiHfNbVZr)$_{1-x}$Al$_x$N was calculated as

$$E_f(x) = E_{total} - \sum_i x_i E_{total}^i, \quad (S.1)$$

where $E_{total}$ is a total energy of a composition at concentrations $x$ modeled as quasi-randomly arranged atoms[8] in B1-NaCl structure, as computed with density-functional theory (DFT), $E_{total}^i$ is the total energy of potential decomposition products, with corresponding weightings $x_i$ to form a considered composition of interest – compositions in Ti-Hf-Nb-V-Zr-Al-N system, including binary and ternary compositions, respectively (as illustrated in Figure 1a), reported previously[9] – computed at the same level of approximations with DFT as implemented elsewhere[10,11]: kinetic energy cutoff $E_{cut}$ is 650 eV, k-spacing in Brillouin zone is 0.5, energy convergence threshold in self-consistent electronic calculations is 10$^{-4}$ eV, ionic geometry relaxation is performed until the 10$^{-3}$ eV threshold is reached.

To account for temperature effects a configurational entropy term is added to the expression $E_f$ in Eq. S.1:

$$E_f = E_f - TS, \quad (S.2)$$

where

$$S = k_B \sum_i x_i \ln x_i \quad (S.3)$$

is configurational entropy, for calculation of which all atomic sites on the metal sublattice in B1-NaCl structure are considered equally accessible for all atomic species in compositions.

# Supplementary Notes

**Supplementary Note 1**

Many studies on high-entropy ceramics focus on maximizing configurational entropy (using equimolar ratios of metal elements) for greater single-phase stabilization[12,13] of single-phase solid solution phase; differing combinations and increasing the number of elements has achieved properties often highly surpassing those of the constituent conventional ceramic components[13,14].

Although some ceramic systems demonstrate high-temperature stability[15,16], recent theoretical studies have shown that all 126 equimolar high-entropy nitrides studied in[17] are metastable with respect to all corresponding equimolar lower-entropy nitride phases that were confirmed experimentally and also proved in other works[16]. Therefore, we propose that exploiting the metastability of high-entropy ceramics can offer additional phase space, resulting in multiple advantages for tuning beneficial properties.

**Supplementary Note 2**

The local lattice disorder inherent in high-entropy ceramics arising from many different sized and chemically distinct elements and fluctuations in the local chemical environment influences the movement of diffusing species, defects, and dislocations[18,19]. Therefore, the high-entropy concept provides ample room for further well-targeted improvement of high-temperature mechanical properties[20]. While mechanical properties are enhanced by severe lattice distortion (solid-solution strengthening), nanograin microstructure impeding dislocation motion (Hall-Petch strengthening), and by changes in and/or elimination of lattice slip systems[21], the inhibition of grain coarsening typical for high-entropy ceramics at elevated temperatures allows to retain mechanical properties[16]. Theoretical calculations demonstrate that high-entropy ceramics show a slow reduction in tensile and shear strength with temperature[22]. Although only a few studies report the mechanical behavior of high-entropy ceramics at elevated temperatures[20,23,24], they demonstrate the exceptional potential of high-entropy ceramics to become new high-temperature materials[25]. However, most rare reports in the literature deal with *post-mortem* mechanical characterization of high-entropy ceramics, and their high-temperature mechanical properties under real operating conditions still lack detailed characterizations[19,26,27]. The possible reason for the latter is that most high-entropy ceramics are synthesized in thin film form. Owing to highly non-equilibrium conditions, physical vapor deposition (PVD) is typically the most preferential method for metastable high-entropy ceramic synthesis. Characterization of mechanical properties of thin films at high temperatures has not been feasible due to the limited volume available and related obstacles. However, recent advancements in thermomechanical testing methods for micro/nano-scale materials allow us to overcome the critical challenges associated with temperature control during material characterization, specimen size, and resolution of measurements[28].

**Supplementary Note 3**

In Fig. S10, the low elastic loading modulus measured at 800 or 900 °C, depending on composition, is arrowed; the lowest modulus is consistently associated with testing at 1000 °C. A measured relatively low modulus is despite the sensor drift corrections, temperature-dependent loading frame modulus correction, Fig. S15, and sink-in correction accounting for the minor (~10%) thermal reduction in elastic modulus of the (0001) sapphire substrate by 1000 °C were considered in prior. Initial suspicion of plastic deformation of the sapphire substrate at high temperature was proven incorrect: a cross-sectional lift out of a hexanary Al-high pillar compressed at 1000 °C imaged by TEM does not reveal either deviation of the film-substrate interface, nor evidence of remaining crystal defects in the sapphire (dislocations or twins), Fig. 4b. According to previous studies, fracture of sapphire at this temperature is preceded by deformation twinning[5]; hence as neither occurs here, this indicates that load applied to the micropillar is effectively distributed in the ~2 µm film thickness below the pillars.

To further elucidate this phenomenon, a TEM analysis of a central cross-section of the diamond punch itself was carried out. According to the SEM imaging, it is evident that imprints of the tops of the micropillars are left in the punch surface after testing up to 1000 °C, Fig. S11. Extensive chemical reactivity between punch and high-entropy refractory nitride pillars was observed, resulting in consumption of the punch and the formation of mechanically weaker oxide phases[29], as well as carbides, oxy-carbides and a carbo-nitride several hundred nanometres thick in total, Fig. S12. Specifically, we identify on the punch surface: $AlO_x$ particles, $(Zr, Hf)O_x$ particles, $CN_x$ particles, a thin $TiC_x$ layer, a thick $(Ti, Zr, Hf, V, Nb, Ta, Al, Si)O_xC_y$ layer interfacing the diamond, and a $SiO_x$ outer layer, where Si originates from a separate Si-doped refractory high-entropy nitride series.

In addition, evidence of mechanical deformation of the diamond (stacking faults on at least 4 distinct systems, as well as an elevated density of ordinary dislocations) was found after testing up to 1000 °C, Fig. S11; these faults are most dense near the contact interface between pillar and punch. Although this is not definitive proof of plasticity generated at high temperature – rather than initially present from punch production or related to low-temperature testing – it is consistent in both crystal deformation mechanism and applied stress (7.9 GPa, vs. ~8.5 GPa maximum here) with previous reports of plasticity in 3.5 GPa anvil-confined 5 µm diamond powder [30].

The chemical reactivity and mechanical deformation of the diamond punch are expected to have occurred most readily at the highest test temperatures, according to straightforward kinetic arguments, and provide pathways for additional punch displacement to be achieved by a given load in the applied test conditions.

**Supplementary Note 4**

W-containing, heptanary nitride: TEM imaging and analysis following micro-compression at 1000 °C revealed extensive oxygen inward diffusion, Fig. S16, manifested as a high oxygen content along columnar grain boundaries in the upper third of the film, Fig. S17, and a ladder-like structure within the grains there, with the horizontal oxygen-rich rungs coincident with locally increased Hf and Zr atomic contents. Diffraction from the low volume of these additional phases was not captured in film's upper half's SAED. Grain-scale segregation of W remains, along with atom-scale segregation.

**Supplementary Note 5**

A simple, classic model of shear strength, $\tau_{sum}$, based upon summative strengthening of contributors yields Eq. S.4, is composed as follows: solid solution strengthening of the metallic sub-lattice, $\tau_{SS}$, supplements the temperature-dependent binary nitride lattice resistance, $\tau^*$, Taylor hardening is given by $\tau_G$, the effect of grain size, diameter $d$, on shear strength, $\tau_{h-p}$, supplements the source-strengthening, $\tau_S$, stemming from the small sample size; finally the contribution of short-range ordering across the metallic sub-lattice is quantified by $\tau_{sr}$, and spinodal decomposition yields the contribution $\tau_{sd}$. Conversion to a uniaxial strength, $\sigma$, is achieved via the Taylor factor $M = 3.06$ for untextured face-centered cubic slipping on the close-packed planes.

$$\tau_{\text{sum}} = \sigma/M = \tau^* + \tau_{\text{SS}} + \tau_{\text{G}} + \tau_{\text{S}} + \tau_{\text{h-p}} + \tau_{\text{sr}} + \tau_{\text{sd}} \qquad (S.4)$$

$$\approx \left(1 - \frac{T}{T_c}\right)\left(\tau_0^* + \tau_{\text{SS},0} + \tau_{\text{sr},0}\right) + \alpha bG\sqrt{\rho} + KG\frac{\ln(\bar{\lambda}/b)}{\bar{\lambda}/b} + \frac{1}{M}K_{\text{h-p}}d^{-1/2} +$$

$$k_s(A\eta Y)^{5/3}(\lambda/Gb)^{2/3} \qquad (S.4b)$$

where material parameters are taken as those for the hexanary Al-high composition (lattice parameter: $a_{\text{6-Al-high}}$ = 4.2672 Å here), as measured and employed in calculations in[1]: shear modulus, $G$ = 168 GPa, the ½<110> Burgers vector of slip, $b$ = 3.017 Å, the modulus term $Y = E(1 - \nu)$, with Young's modulus $E$ = 453 GPa and Poisson ratio $\nu$ = 0.209; $k_s = 0.122[\pi(1 - \nu)(1 - 2\nu)]^{2/3}$ is another material elasticity term. Further parameters include the dislocation density, $\rho$, taken as $10^{15}$ m$^{-2}$ here, $K_{\text{h-p}}$ = 12.5 GPa nm$^{-1/2}$ the Hall-Petch constant (taken for TiN[31]), a $d$ = 100 nm grain size, micropillar diameter $D$ = 1800 nm which approximates the average source length $\bar{\lambda}$, and pre-factors $K$ = 0.5 & $\alpha$ = 0.1. The solid solution- and short-range order-strengthened lattice resistance prefactor is approximated here as: $\left(\tau_0^* + \tau_{\text{SS},0} + \tau_{\text{sr},0}\right)$ = 4.5 GPa; $T_c$ is a critical temperature (beyond the temperature range probed here), above which lattice resistance is assumed constant and low. With regards to the spinodal itself, $A = (C - C_0)/3$, is the amplitude of each of the three dimensional composition modulation components on the metallic sub-lattice, giving $A$ = (92 – 51) / 3 = 13.7 at.% here for the Al composition in the hexanary Al-high nitride after annealing at 1200 °C, Fig. 2f, whilst $\eta = a^{-1} \, da/dC$ is the relative lattice parameter change against composition, and the spinodal wavelength of a specific material and annealing condition is $\lambda$. Upon further inspection the equation for the spinodal component simplifies if Vegard's law of proportionality between isostructural solid solutions' compositions and their lattice parameters holds true: hence $\eta \approx a^{-1} \Delta a/\Delta C$ where $\Delta$ denotes a finite change in a value, with the result that $A\eta$, and thereby $\tau_{\text{sd}}$, can be expressed as a function of the amplitude of the change in lattice parameter across the spinodal, without including the explicit value of composition change itself – which cancels between $A$ and $\eta$. It is evident that any real deviations from Vegard's law leading to an increased $da/dC$ will yield additional strengthening. The lattice parameter of cubic AlN is taken as 4.045 Å.

# Supplementary Tables

**Table S1.** Elemental composition of the samples derived from the combined measurements using XPS and ToF-ERDA.

| Sample | Composition (at.%) | | | | | | | | | | | |
|---|---|---|---|---|---|---|---|---|---|---|---|---|
| | N | Al | Ti | Hf | V | Zr | Nb | Ta | W | Ar | O | C |
| (TiVZrNbHf)N | 48.2 | – | 11.8 | 7.6 | 10.5 | 12.2 | 7.8 | – | – | 0.4 | 0.8 | 0.7 |
| (TiVZrNbHf)$_{0.86}$Al$_{0.14}$N | 48 | 6.9 | 10.3 | 6.5 | 9.0 | 10.2 | 6.8 | – | – | 0.4 | 1 | 0.8 |
| (TiVZrNbHf)$_{0.49}$Al$_{0.51}$N | 48.8 | 25.0 | 6.3 | 3.4 | 5.1 | 5.8 | 3.5 | – | – | 0.4 | 0.8 | 0.7 |
| (TiVZrNbHfTa)N | 48.2 | – | 11.5 | 7.6 | 6.9 | 10.4 | 6.5 | 6.7 | – | 0.5 | 0.9 | 0.8 |
| (TiVZrNbHfTaW)N | 45.0 | – | 9.7 | 5.6 | 7.3 | 9.5 | 5.9 | 5.6 | 9.6 | 0.5 | 0.6 | 0.9 |

# Supplementary Figures

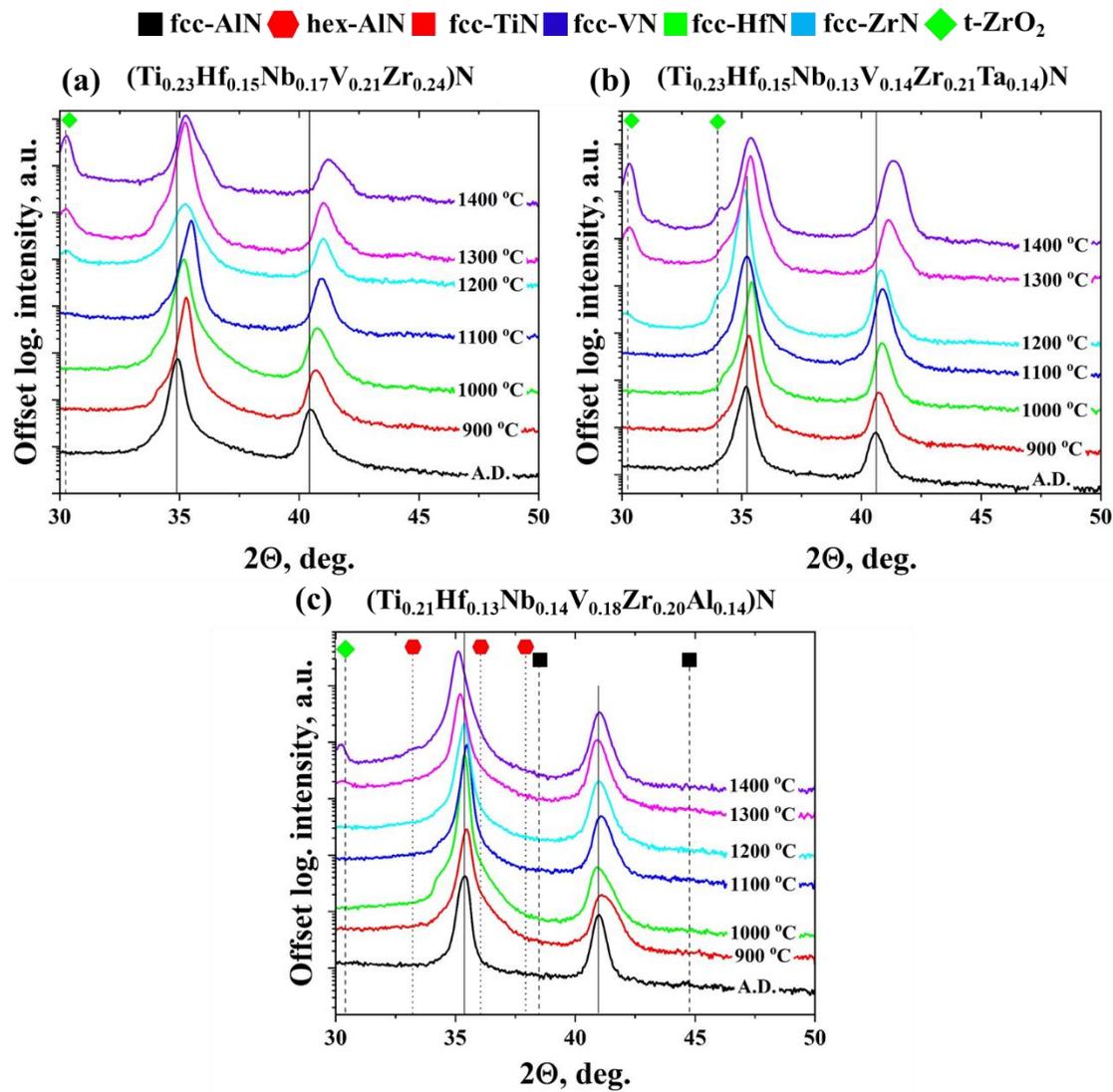

**Fig. S1.** (a-c) XRD patterns for pentanary $(Ti_{0.23}Hf_{0.15}Nb_{0.17}V_{0.21}Zr_{0.24})N$, hexanary $(Ti_{0.23}Hf_{0.15}Nb_{0.13}V_{0.14}Zr_{0.21}Ta_{0.14})N$, and hexanary 'Al-low' $(Ti_{0.21}Hf_{0.13}Nb_{0.14}V_{0.18}Zr_{0.20}Al_{0.14})N$ thin films before and after annealing in vacuum at 900, 1000, 1100, and 1200 °C.

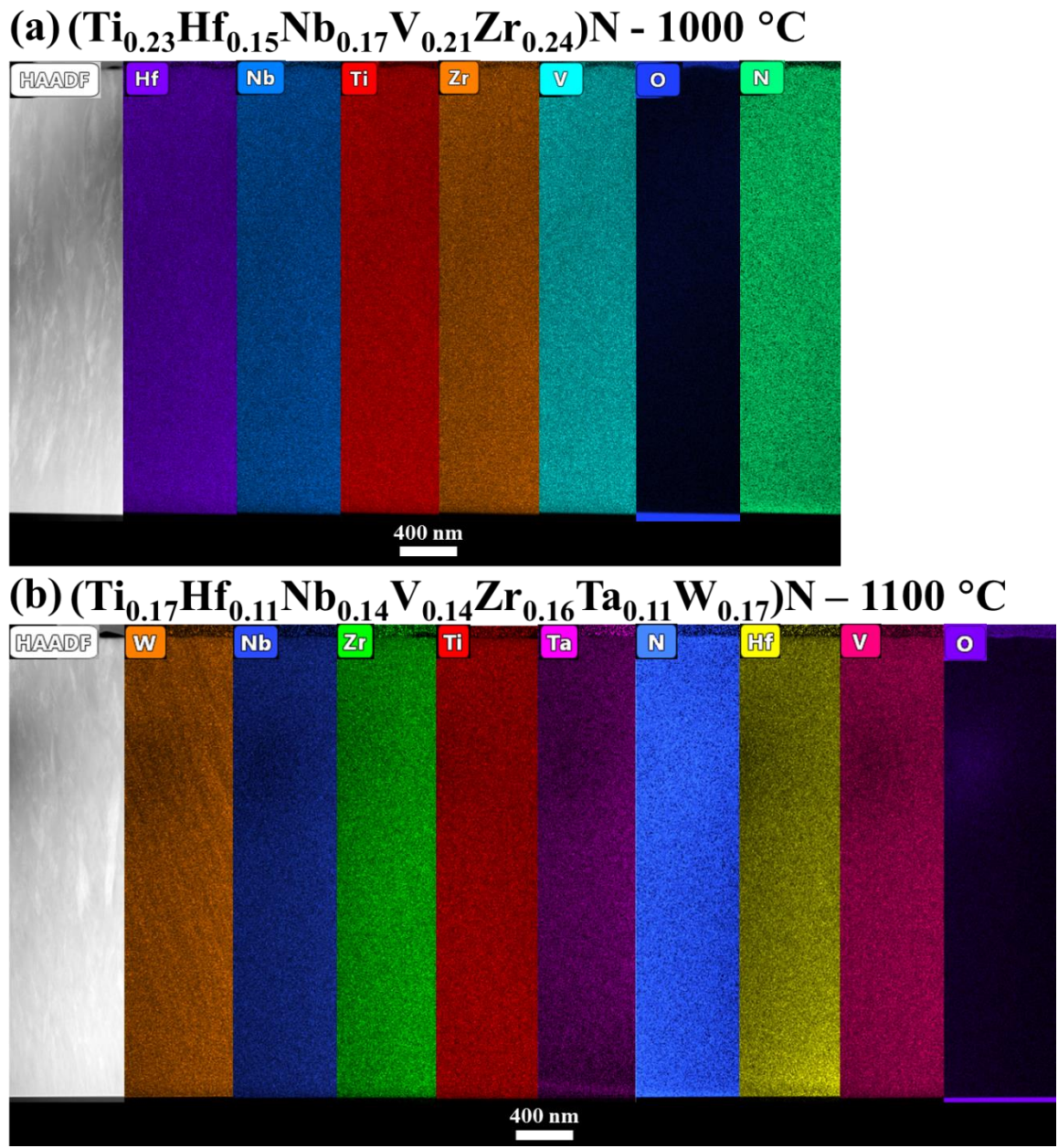

**Fig. S2.** EDX color-coded elemental maps acquired from (a) pentanary nitride annealed at 1000 °C and (b) heptanary nitride annealed at 1100 °C.

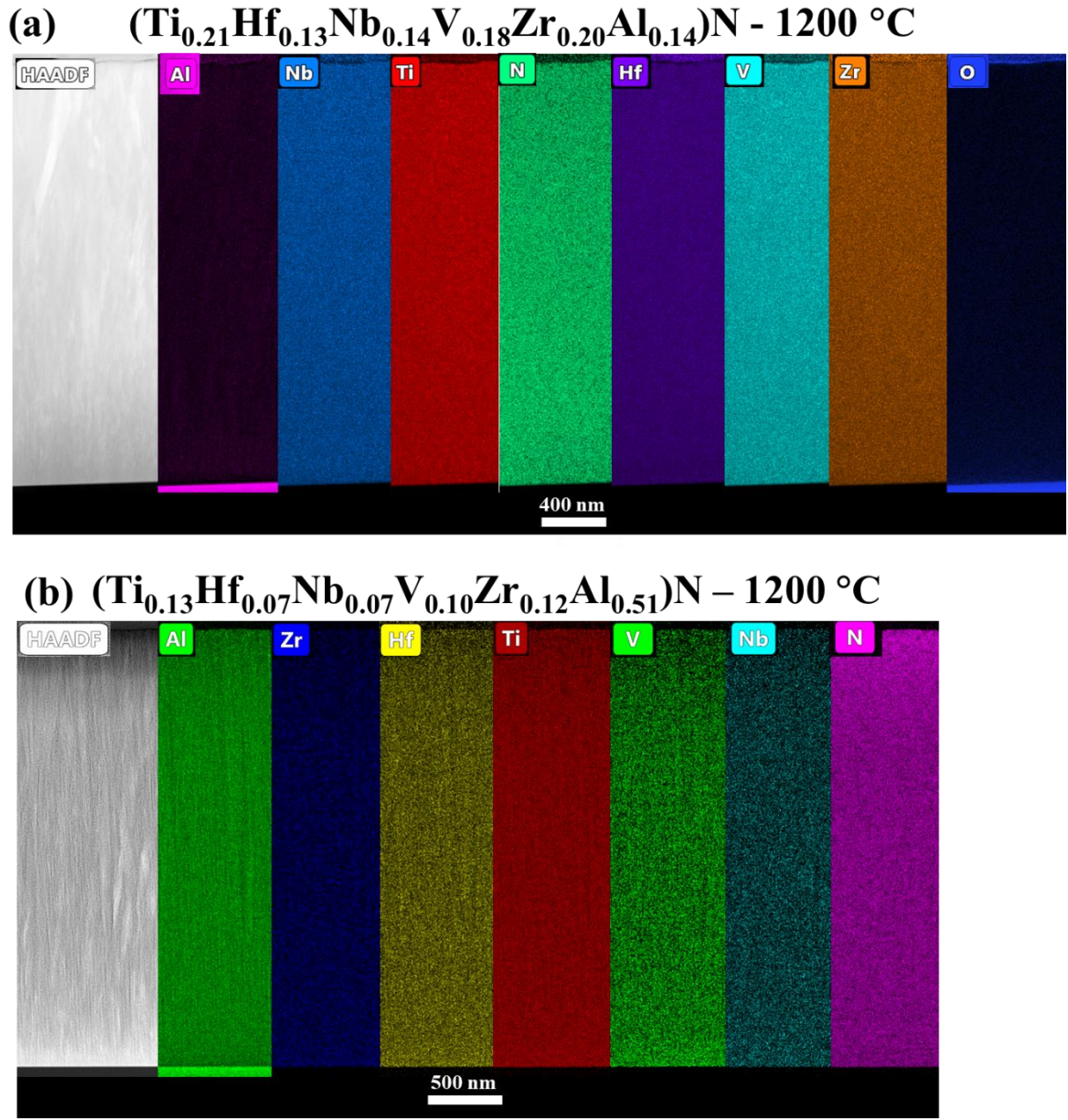

**Fig. S3.** EDX color-coded elemental maps acquired from (a) the hexanary 'Al-low' $(Ti_{0.21}Hf_{0.13}Nb_{0.14}V_{0.18}Zr_{0.20}Al_{0.14})N$ and (b) hexanary 'Al-high' $(Ti_{0.13}Hf_{0.07}Nb_{0.07}V_{0.10}Zr_{0.12}Al_{0.51})N$ films after annealing at 1200 °C.

$(Ti_{0.17}Hf_{0.11}Nb_{0.14}V_{0.14}Zr_{0.16}Ta_{0.11}W_{0.17})N$
− 1100 °C

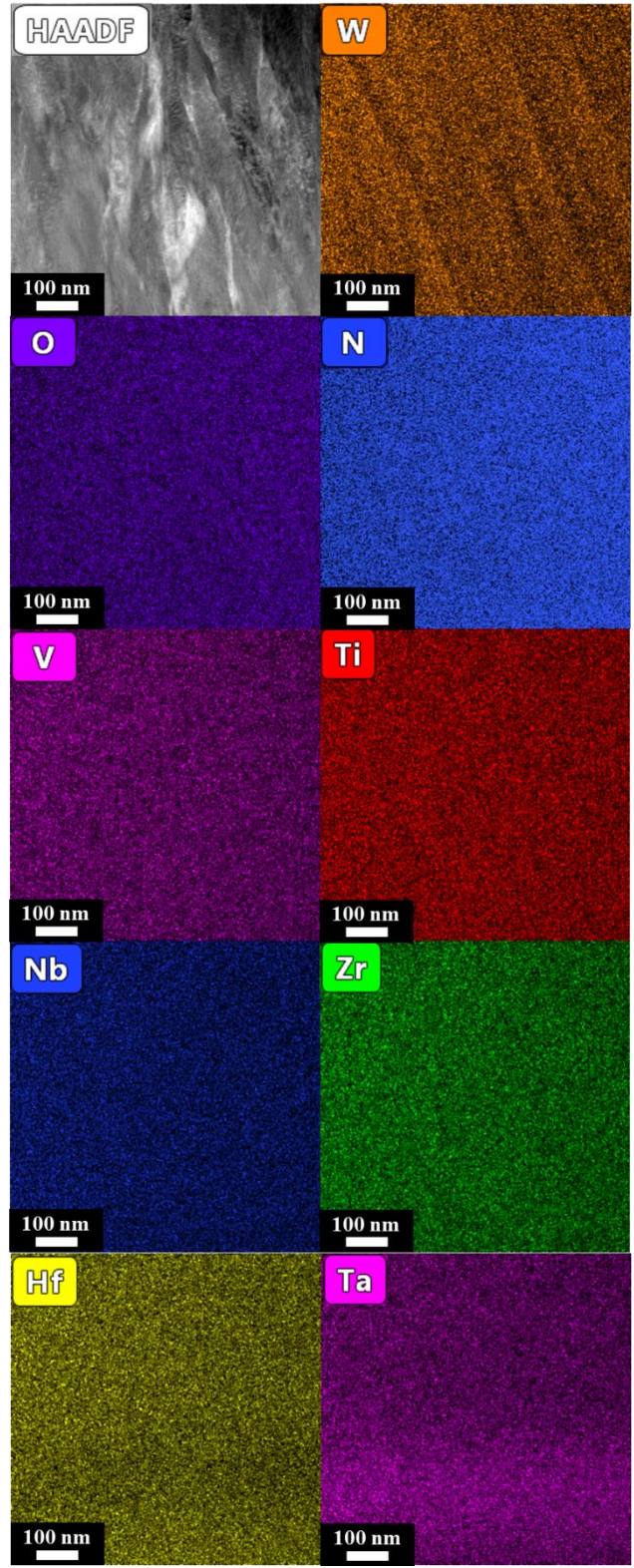

**Fig. S4** Magnified EDX color-coded elemental maps acquired from heptanary nitride annealed at 1100 °C.

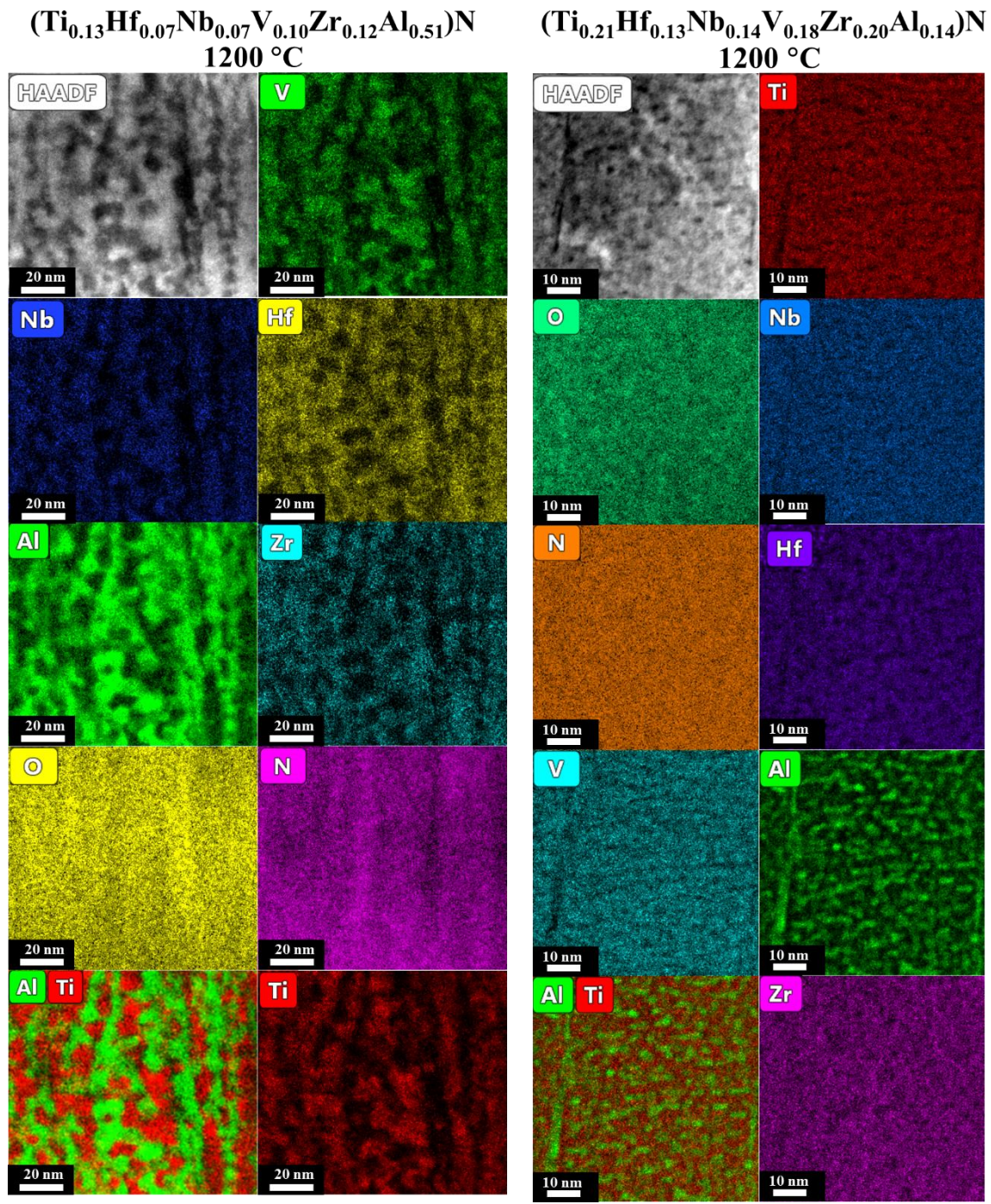

**Fig. S5** Magnified EDX color-coded elemental maps acquired from the hexanary 'Al-high' $(Ti_{0.13}Hf_{0.07}Nb_{0.07}V_{0.10}Zr_{0.12}Al_{0.51})N$ and the hexanary 'Al-low' $(Ti_{0.21}Hf_{0.13}Nb_{0.14}V_{0.18}Zr_{0.20}Al_{0.14})N$ films after annealing at 1200 °C.

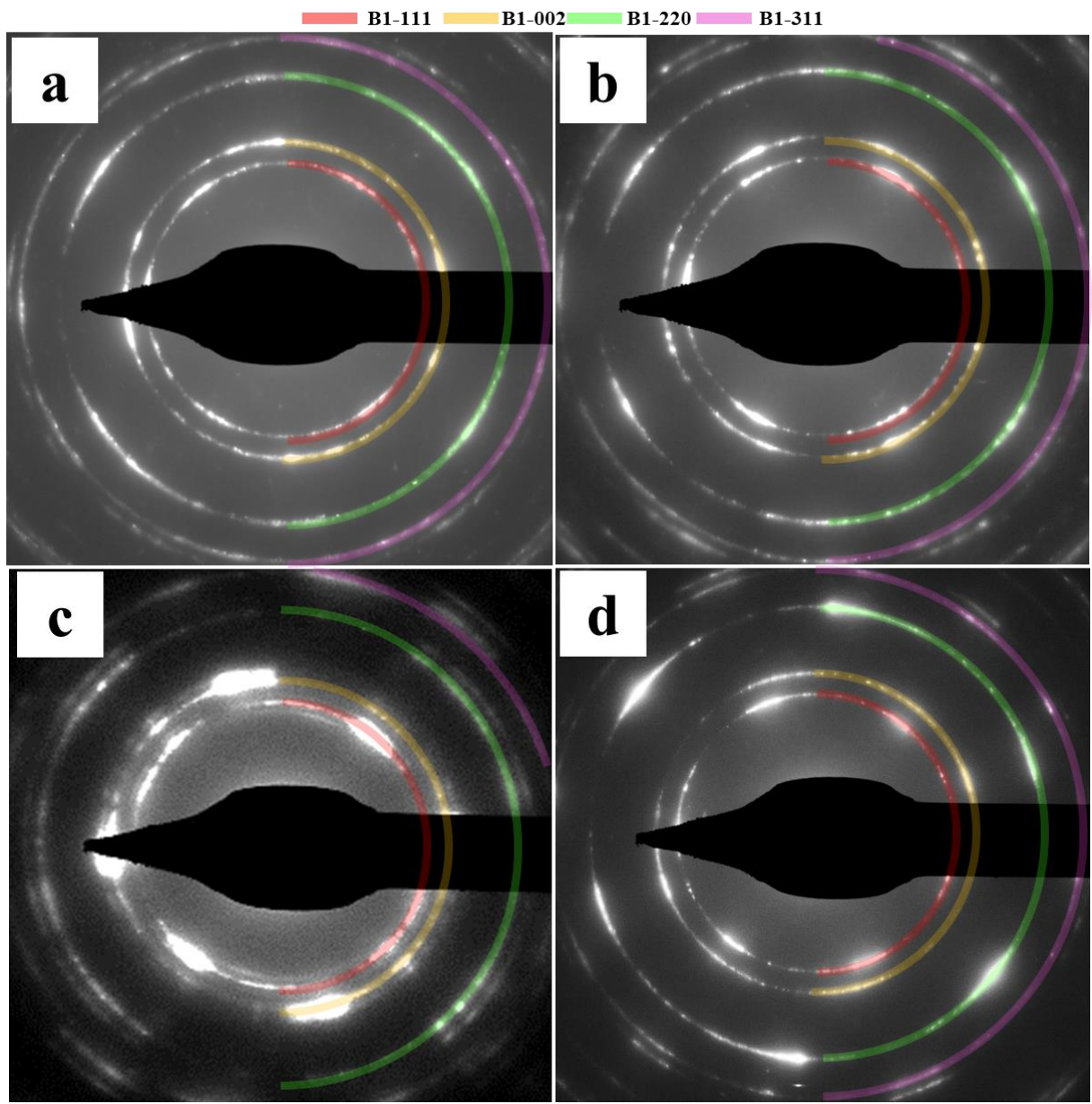

**Fig. S6.** SAED acquired from (a) heptanary nitride annealed at 1100 °C, (b) pentanary nitride annealed at 1000 °C, (c) hexanary 'Al-high' and (d) hexanary 'Al-low' films after annealing at 1200 °C. The film growth direction is horizontal for the SAEDs presented here.

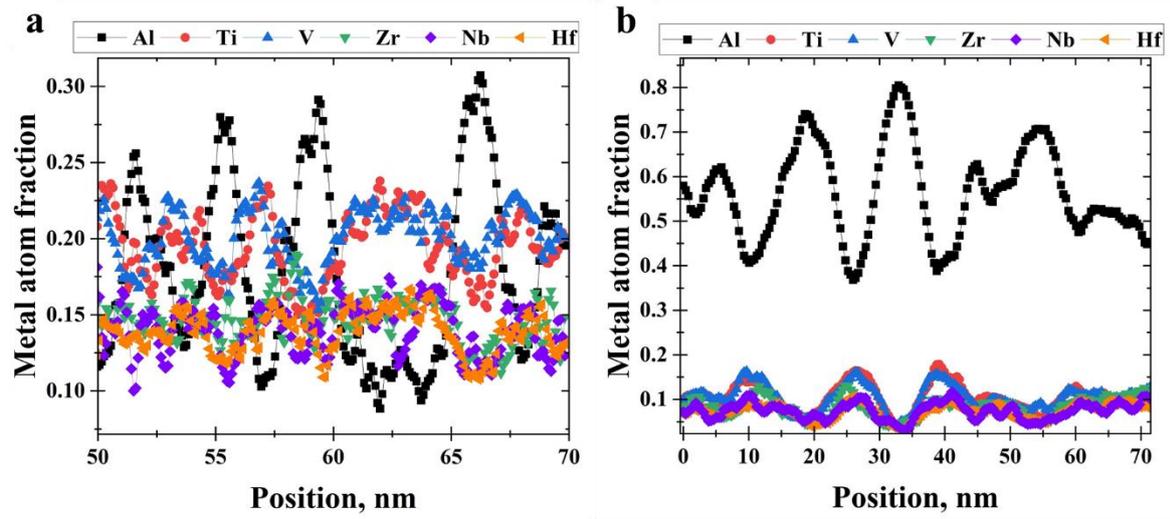

**Fig. S7.** EDS line scans acquired from (a) hexanary 'Al-low' and (b) hexanary 'Al-high' films after annealing at 1200 °C.

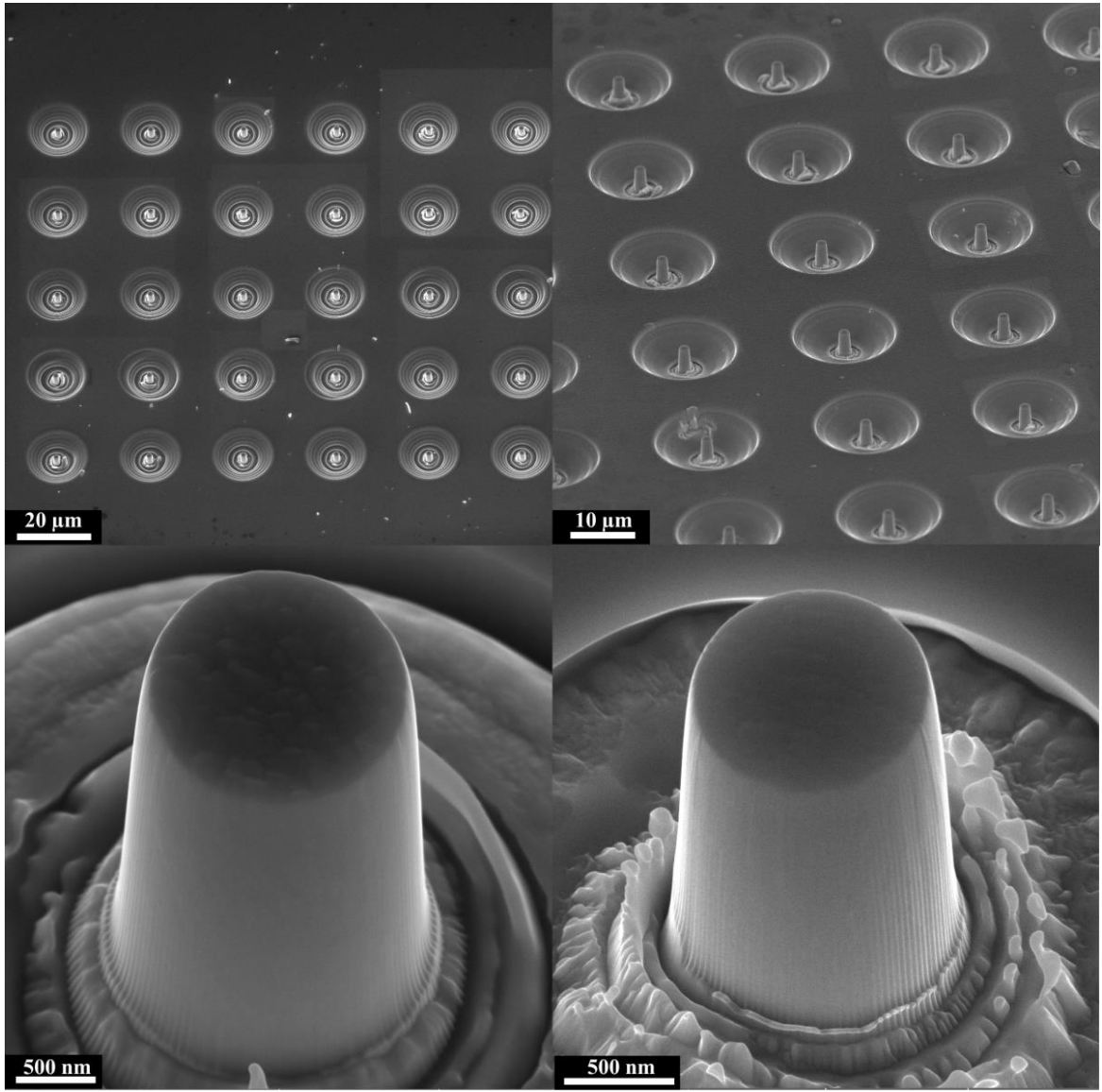

**Fig. S8.** SEM images of representative sets of FIB-milled micro-pillars before microcompression, viewed at 45° tilt angle.

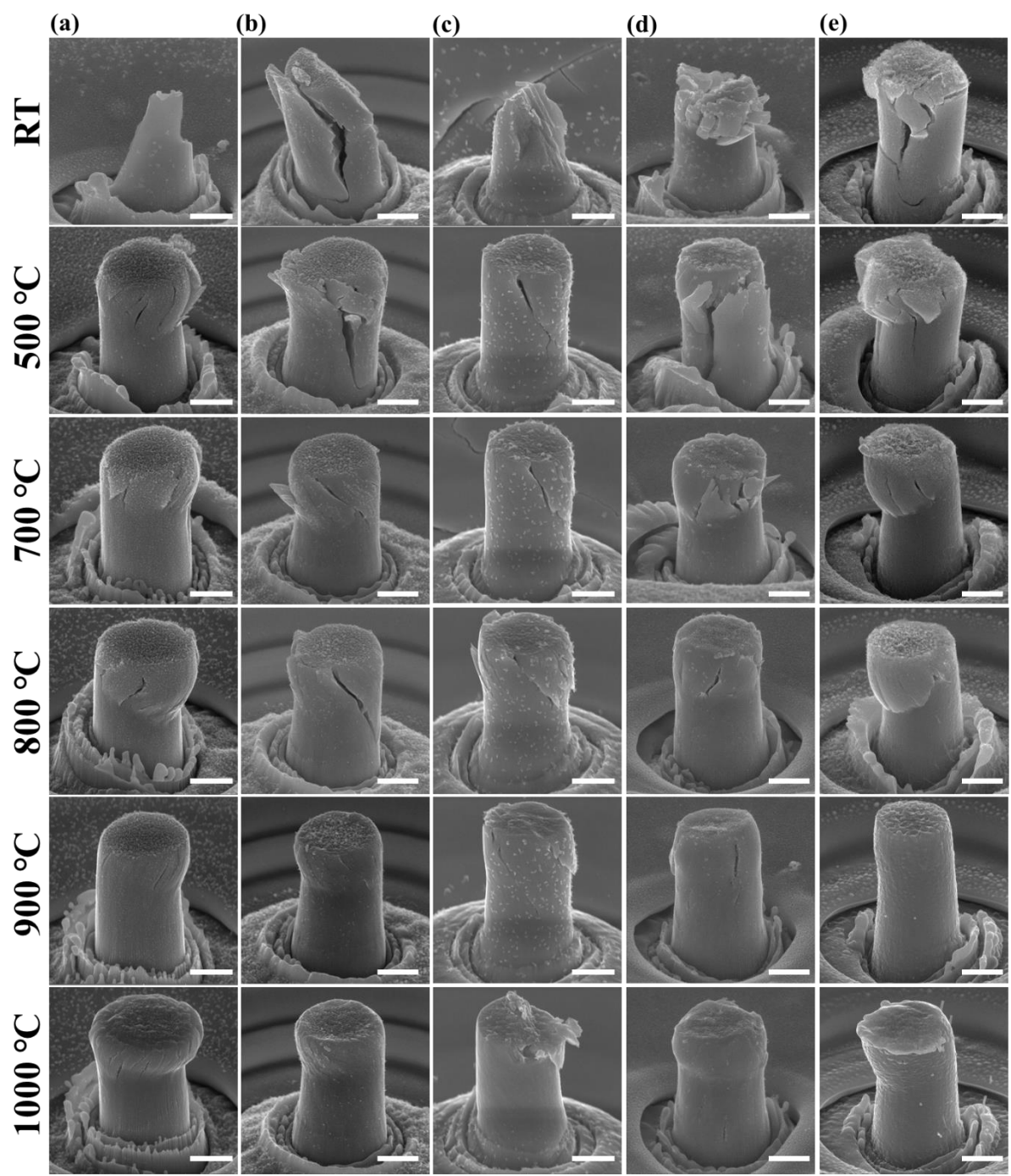

**Fig. S9.** SEM images of micro-pillars deformation and fracture for (a) pentanary, (b) hexanary, (c) heptanary, (d) hexanary Al-low and (e) Al-high systems after compression at room temperature (RT), 500 °C, 700 °C, 800 °C, 900 °C, and 1000 °C. All scale bars are 1 μm.

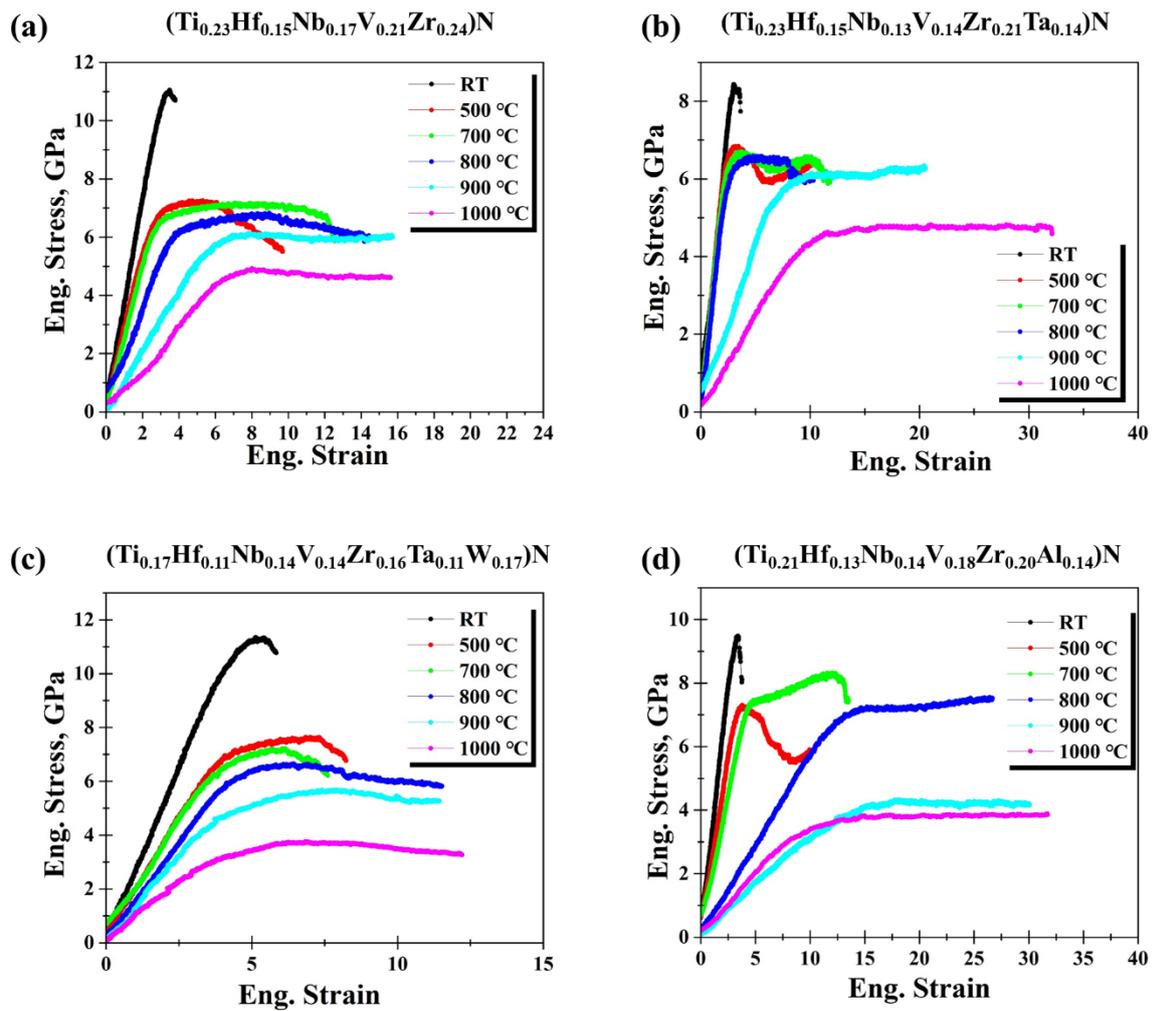

**Fig. S10.** A set of stress-strain curves for (a) pentanary $(Ti_{0.23}Hf_{0.15}Nb_{0.17}V_{0.21}Zr_{0.24})N$, (b) hexanary $(Ti_{0.23}Hf_{0.15}Nb_{0.13}V_{0.14}Zr_{0.21}Ta_{0.14})N$, and (c) heptanary $(Ti_{0.17}Hf_{0.11}Nb_{0.14}V_{0.14}Zr_{0.16}Ta_{0.11}W_{0.17})N$, (d) the hexanary Al-low $(Ti_{0.21}Hf_{0.13}Nb_{0.14}V_{0.18}Zr_{0.20}Al_{0.14})N$ systems compressed at RT, 500 °C, 700 °C, 800 °C, 900 °C, and 1000 °C. For statistical accuracy, at least 4-5 pillars are compressed for each film composition at each test temperature, but only representative stress-strain curves are shown here.

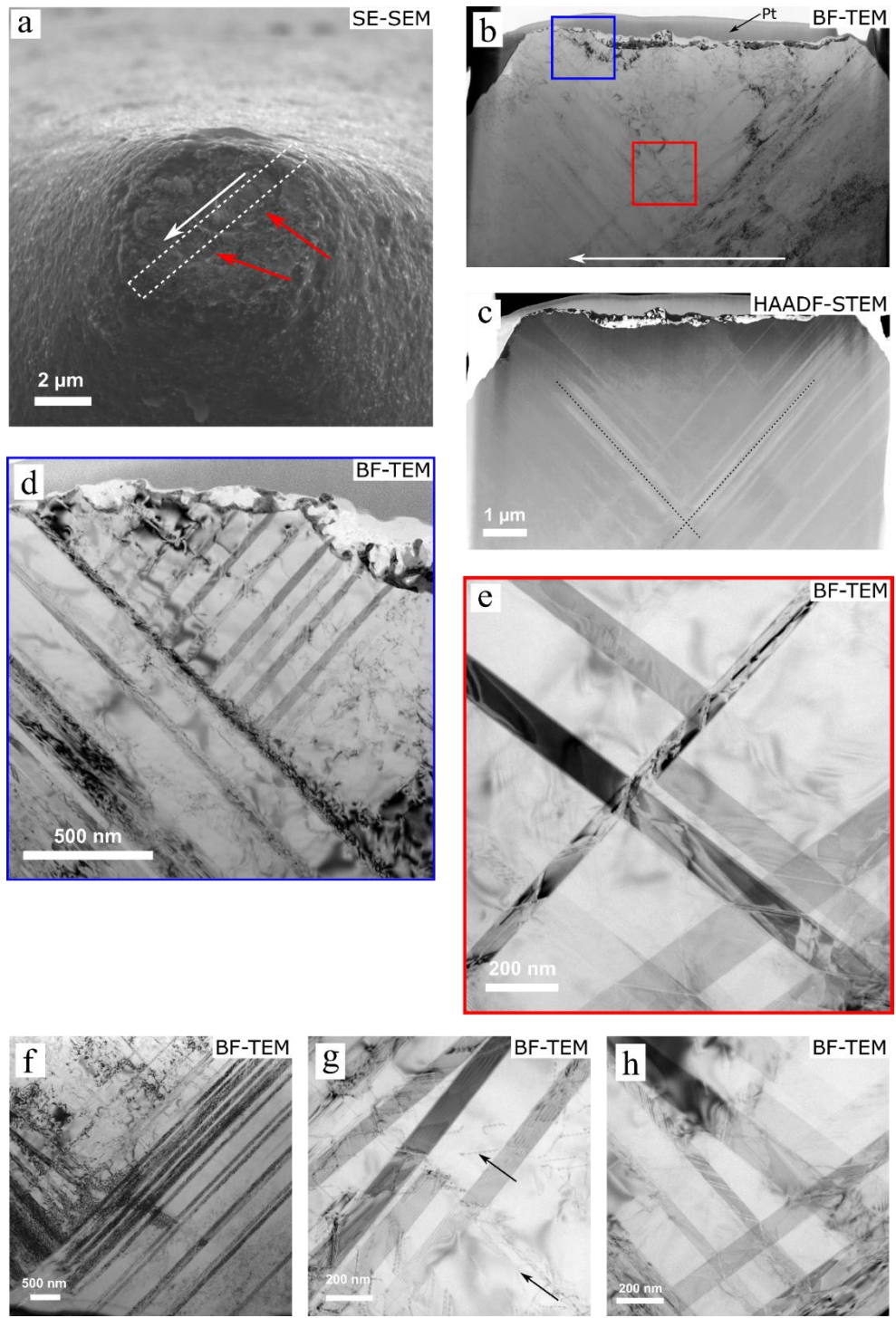

**Fig. S11.** (a) SEM image of the diamond punch after compression up to 1000 °C with pillar imprints arrowed in red and cross-section lift-out location indicated; (b, c) BF-TEM and STEM-HAADF overview images of the punch cross-section, oriented by the white arrows in (a, b), with significant intersecting slip traces indicated (dotted black lines in (c)). BF-TEM images of the crystal deformation structures (d) in the region near the micropillar contact surface, indicated in (b) and (e - h) at the intersection of significant stacking fault families, where the stacking fault density is high. Black arrows in (g) indicate ordinary dislocations.

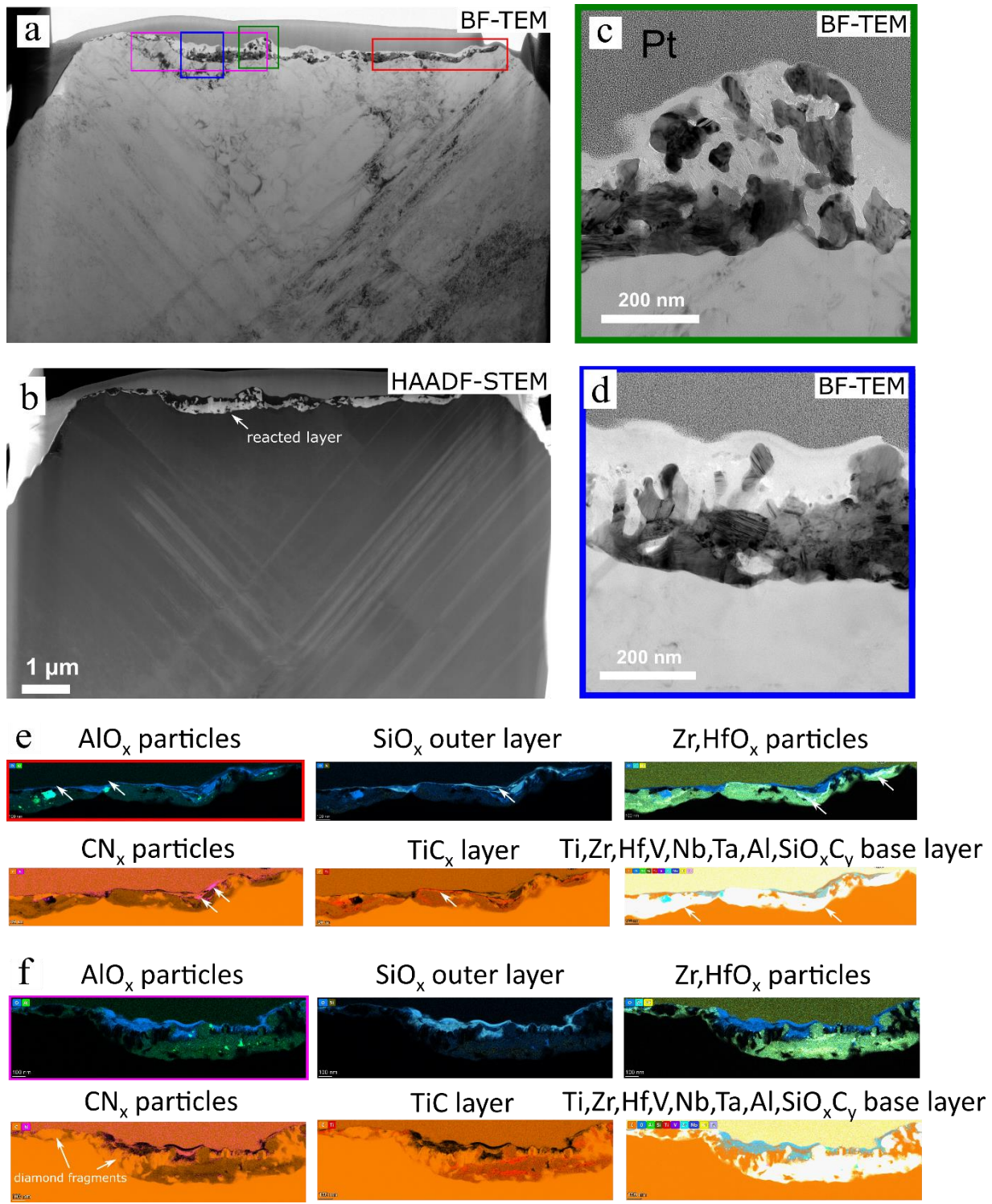

**Fig. S12.** (a, b) BF-TEM and STEM-HAADF overview images of the punch cross-section, (c, d) High magnification BF-TEM images of the reacted layer, and (e, f) STEM-EDX analysis of the reaction products, indicated by white arrows in each case, observed in the cross-section of the diamond punch after compression up to 1000 °C, taken from the regions highlighted in (a).

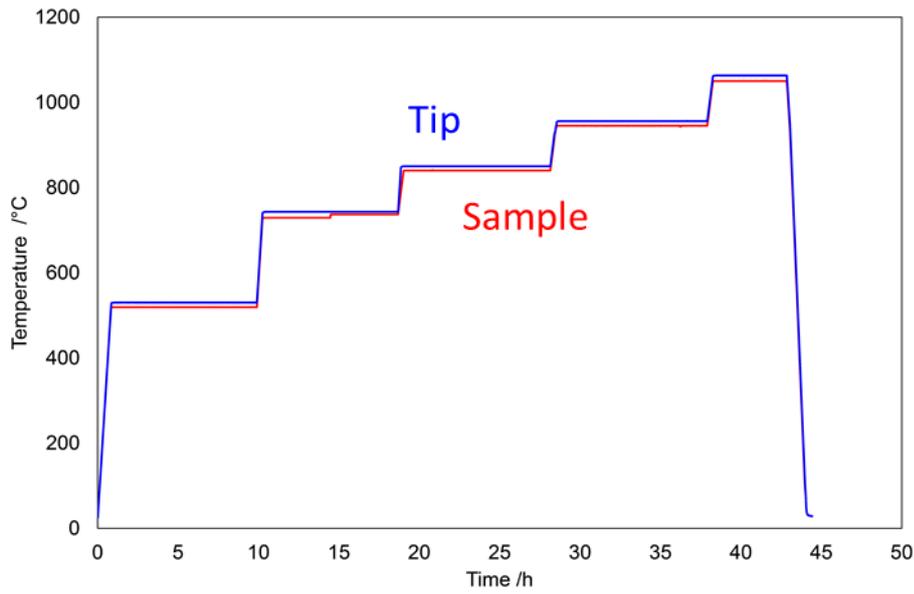

**Fig. S13.** Temperature profile of a representative high-temperature microcompression experiment: measured temperatures of tip and sample side thermocouples.

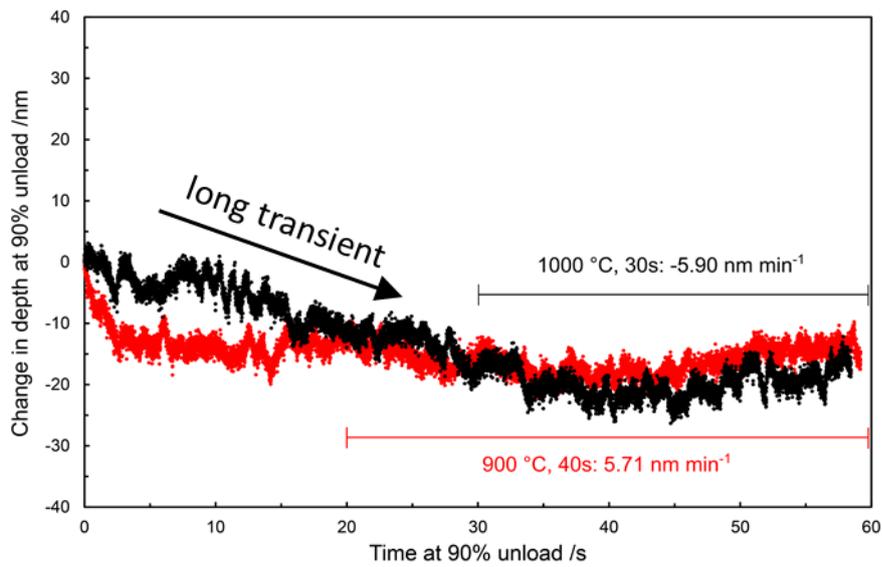

**Fig. S14.** Displacement drift curves at constant load (upon 90% unloading) for temperature matching at 900 and 1000 °C.

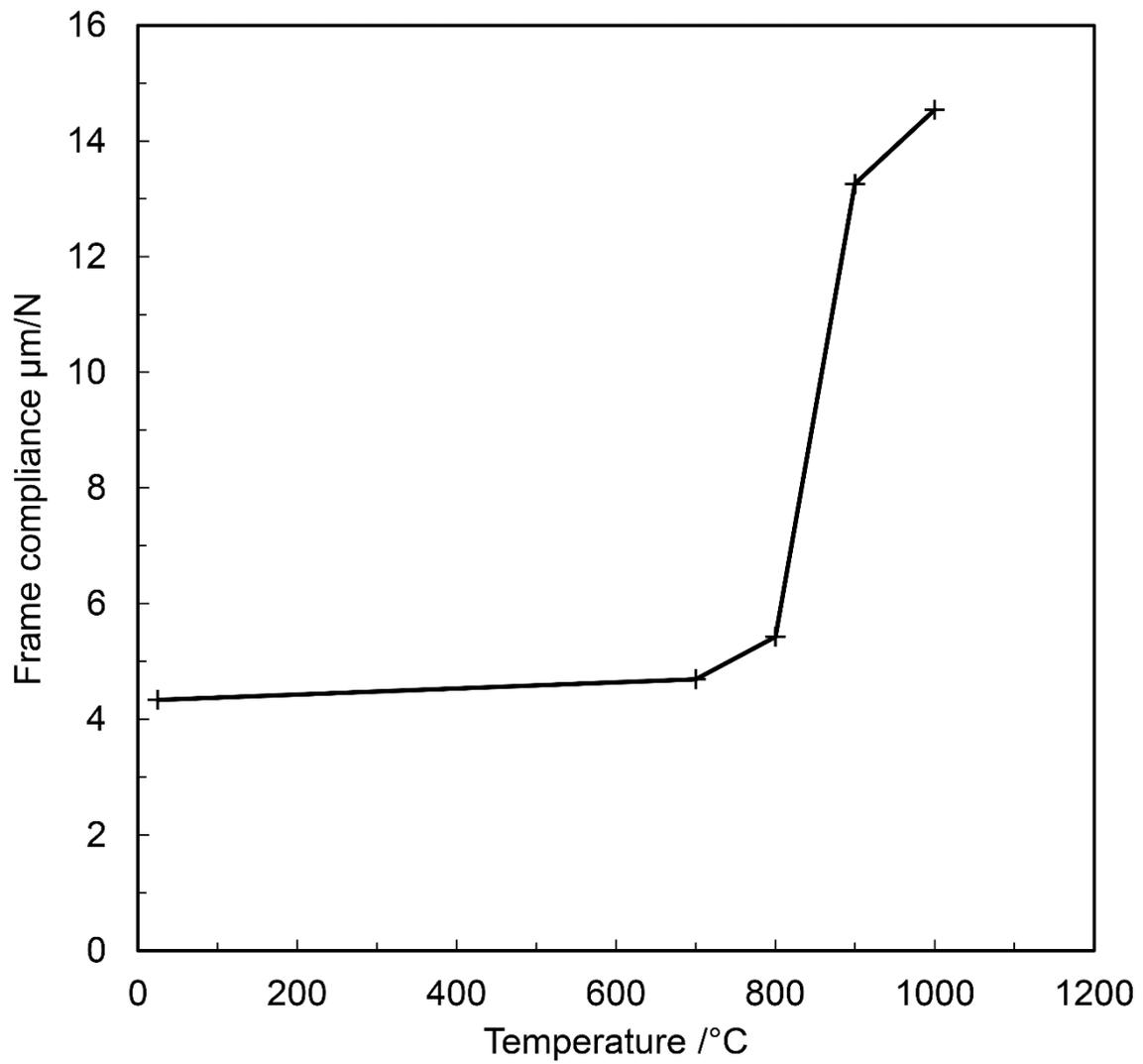

**Fig. S15.** Rig compliance of the high-temperature nanoindentation setup against test temperature measured using a conductive diamond Berkovich on (0001) sapphire.

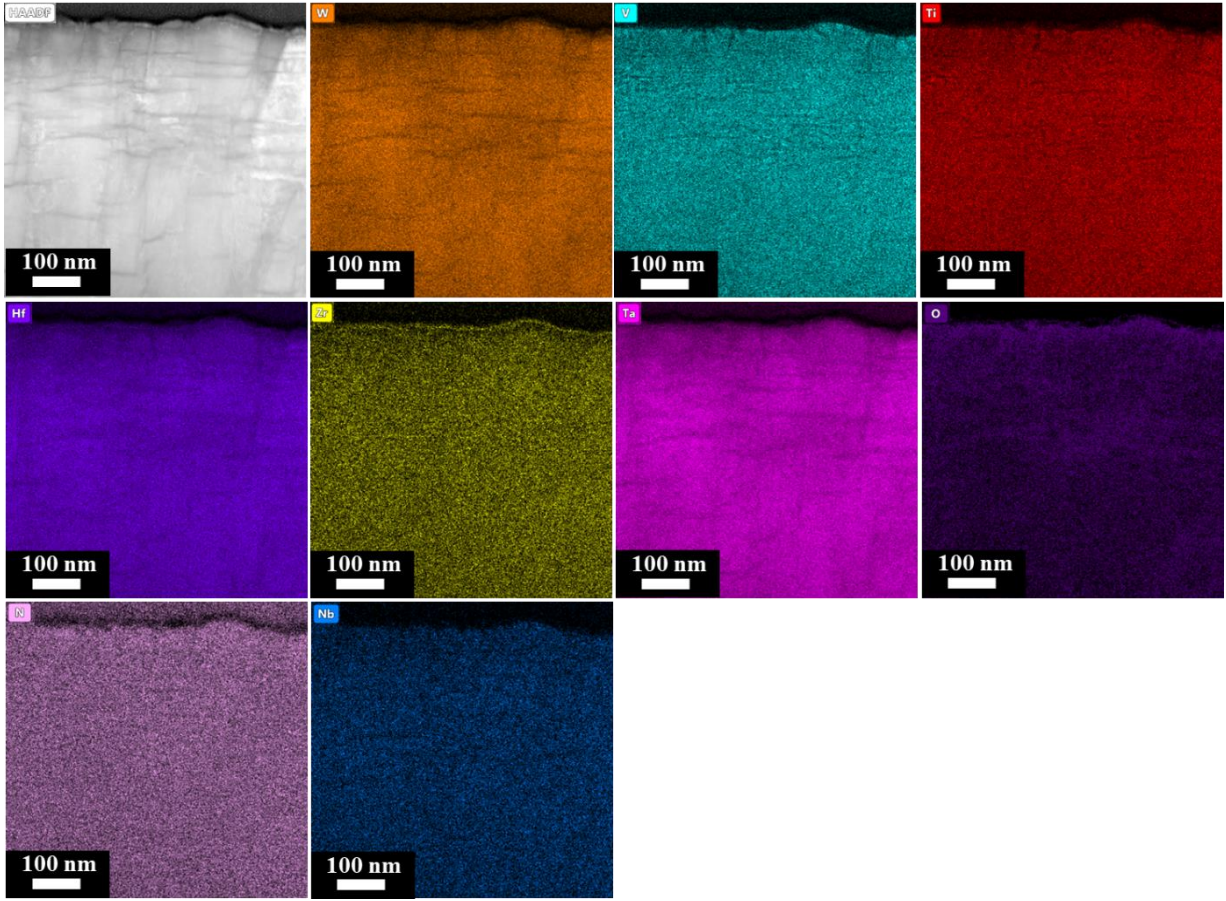

**Fig. S16.** Magnified STEM-EDX color-coded elemental maps acquired from the near-surface area of the heptanary nitride film after micropillar compression tests at 1000 °C.

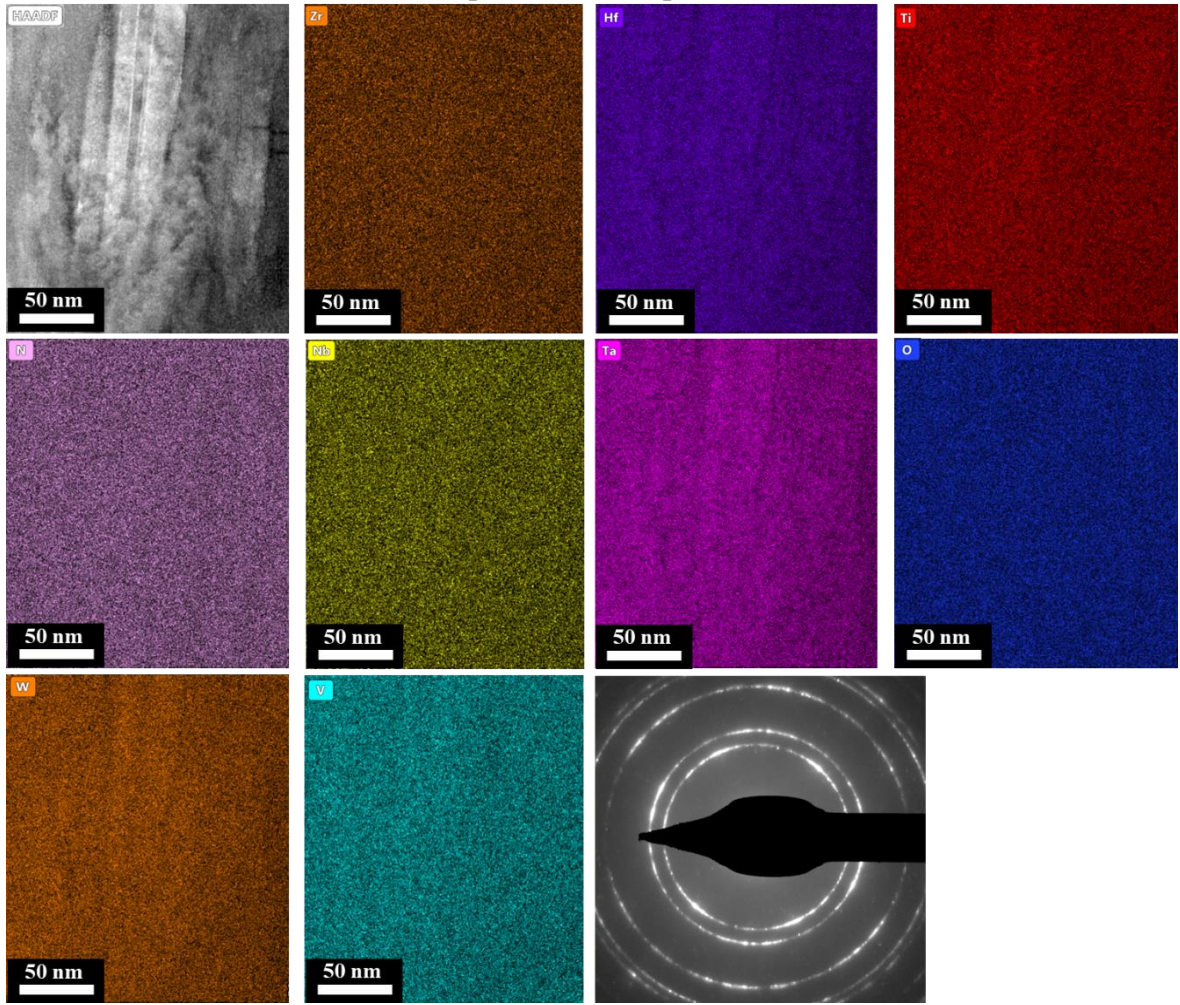

**Fig. S17.** Magnified STEM-EDX color-coded elemental maps acquired from the mid-depth of the heptanary nitride film after micropillar compression tests at 1000 °C and corresponding SAED.

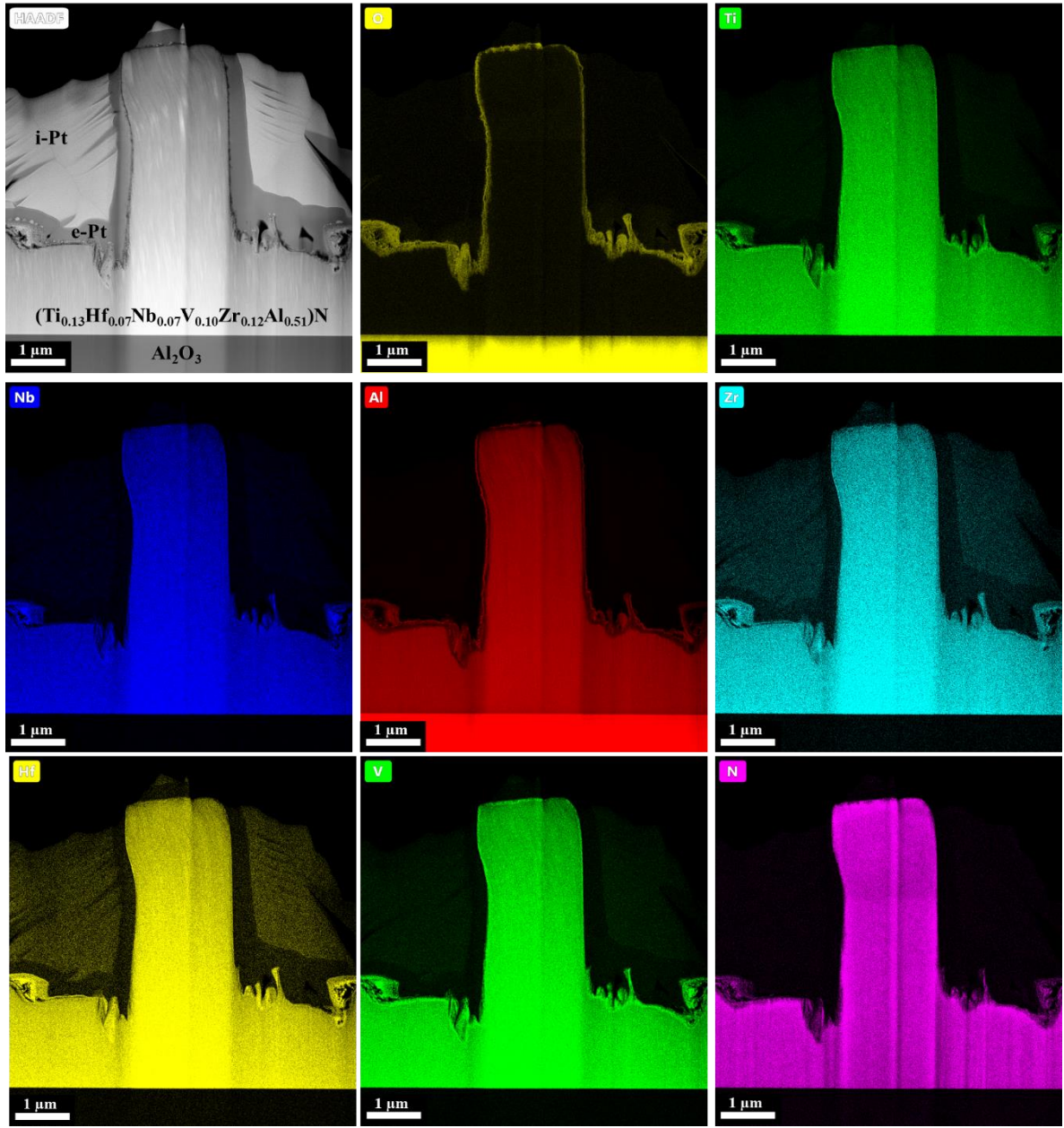

**Fig. S18.** STEM-EDX color-coded elemental maps acquired from the hexanary Al-high pillar after compression at 1000 °C.

# (Ti$_{0.13}$Hf$_{0.07}$Nb$_{0.07}$V$_{0.10}$Zr$_{0.12}$Al$_{0.51}$)N
## 1000 °C-compressed pillar

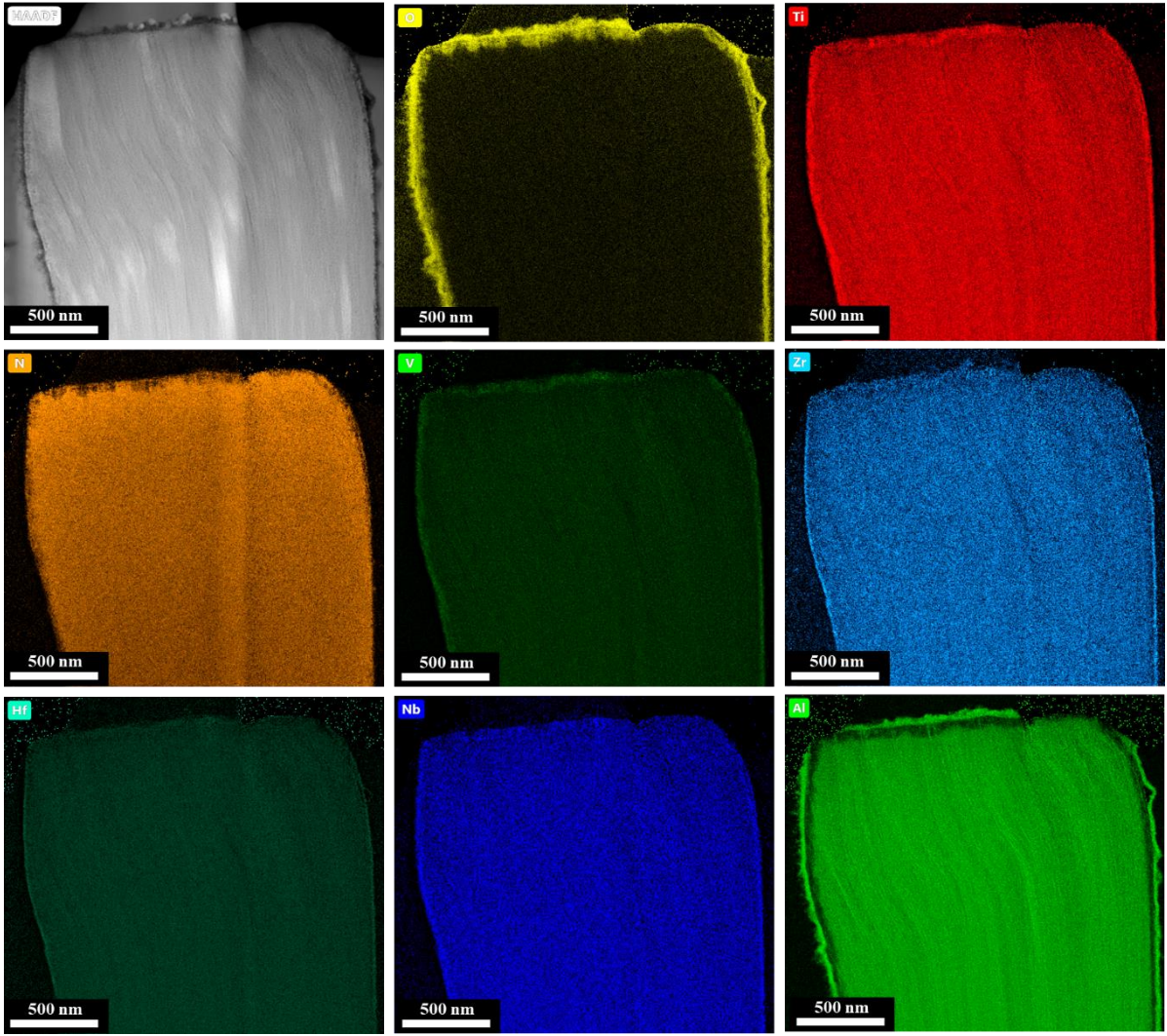

**Fig. S19.** STEM-EDX color-coded elemental maps acquired from the upper part of the hexanary Al-high pillar after compression at 1000 °C.

# Additional References